\documentstyle[aps,twoside]{revtex}
\oddsidemargin=\evensidemargin

\begin{document}
\twocolumn[\hsize\textwidth\columnwidth\hsize\csname@twocolumnfalse%
\endcsname
\draft

\title{Derivation of the Effective Chiral Lagrangian for Pseudoscalar Mesons
from QCD}

\author{Qing Wang$^{a,b}$, Yu-Ping Kuang$^{b,a}$,
Xue-Lei Wang$^{a,c}$, Ming Xiao$^a$}

\address{$^a$Department of Physics, Tsinghua University, Beijing 100084, China
\footnote{Mailing address}\\
$^b$China Center of Advanced Science and Technology (World Laboratory),
P.O.Box 8730, Beijing 100080, China \\
$^c$Department of Physics, Henan Normal University, Xinxiang 453002, China}
\date{TUIMP-TH-99/100, Feb 28,1999}

\maketitle

\begin{abstract}
We formally derive the chiral Lagrangian for low lying 
pseudoscalar mesons from the first principles of QCD considering the
contributions from the normal part of the theory without taking 
approximation. The derivation is based on the standard generating functional 
of QCD in the path integral formalism. The gluon-field integration is formally 
carried out by expressing the result in terms of physical Green's functions of 
the gluon. To integrate over the quark-field, we introduce a bilocal auxiliary
field $\Phi(x,y)$ representing the mesons. We then develop a consistent way of
extracting the local pseudoscalar degree of freedom $U(x)$ in $\Phi(x,y)$
and integrating out the rest degrees of freedom such
that the complete pseudoscalar degree of freedom resides in $U(x)$. With
certain techniques, we work out the explicit $U(x)$-dependnce of the effective 
action up to the $p^4$-terms in the momentum expansion, which leads to the
desired chiral Lagrangian in which all the coefficients contributed
from the normal part of the theory are expressed in terms of certain
quark Green's functions in QCD. Together with the exsisting QCD formulae
for the anomaly contributions, the present results leads to the complete 
effective chiral Lagrangian for pseudoscalar mesons. The final result can be 
regarded as the fundamental QCD definition of the coefficients in the chiral 
Lagrangian. The relation between the present QCD definition of the $p^2$-order 
coefficient $F^2_0$ and the well-known appoximate result given by Pagels and 
Stokar is discussed.
\end{abstract}

\bigskip
PACS number(s):  12.39.Fe, 11.30.Rd, 12.38.Aw, 12.38.Lg,

\vspace{0.5cm}
]
\null\vspace{0.2cm} 
\begin{center}
{\bf I. INTRODUCTION}
\end{center}

The study of low energy hadron physics in QCD is a long standing difficult
problem due to its nonperturbative nature. For low lying pseudoscalar mesons, 
a widely used approach is the theory of chiral Lagrangian based on 
the momentum expansion and the consideration of the global symmetry of the 
system without dealing with the nonperturbative dynamics of QCD 
\cite{weinberg}\cite{GL}. In the chiral Lagrangian approach, the coefficients 
in the Lagrangian are all unknown phenomenological parameters which are 
determined by experimental inputs. The number of the unknown parameters 
increases rapidly with the increase of the precision in the momentum expansion.
For example, the chiral Lagrangian for pseudoscaler mesons with three flavors 
up to the $p^4$-terms given by Gasser and Leutwyler \cite{GL} contains 14 
unknown coefficients. When the $p^6$-terms are taken into account, there are 
143 additional unknown coefficients \cite{FS}.

 This kind of approach has also been applied to the electroweak theory 
\cite{AW} for studying the probe of the electroweak symmetry breaking 
mechanism \cite{BDV,HKY}. Since parity and CP are not conserved in the 
electroweak theory, there are even more unknown coefficients in the 
electroweak chiral Lagrangian than in the case of QCD \cite{AW}. So far, this 
kind of study is of the level of finding out suitable processes at 
future high energy colliders to determine the unknown coefficients in the 
electroweak chiral Lagrangian and investigating to what precision the 
determination can be. The relation between the coefficients in the electroweak 
chiral Lagrangian and the underlying model of the electroweak symmetry 
breaking mechanism is yet not known except for some very simple models 
\cite{BDV}.

Further study on understanding the relation between the chiral 
Lagrangian coefficients and the underlying dynamical theory will be very 
helpful both in QCD and in the electroweak theory for reducing the number of 
independent unknown parameters which makes the theory more predictive.
There are papers studying approximate formulae for the chiral Lagrangian 
coefficients based on certain dynamical ansatz \cite{Holdom}, but the approach
is not completely from the first principles of the underlying theory. Attempts 
to build closer relations between the chiral Lagrangian and the long distance 
piece of the underlying theory of QCD by considering the anomaly contributions 
with certain approximation also exist \cite{derive,Espriu}. However, several 
aspects of it imply that such kind of approach needs improvement, e.g. (a) the 
theory does not include spontaneous chiral symmetry breaking, and the chiral 
symmetry breaking scale is put in by hand; (b) without putting in the chiral 
symmetry breaking scale, the obtained pion decay constant $F_\pi$ is 
proportional to an imposed very low ($\sim 320$ MeV) momentum cut-off on the 
underlying theory of QCD; (c) the positivity of $F^2_\pi$ depends on a careful 
choice of the regularization scheme. The approach in Ref.\cite{Simic}
does not contain the above problems. But in Ref.\cite{Simic}, the
approximation of large-$N_c$ limit is taken from the beginning and 
the approximation of picking up only the local scalar and pseudoscalar pieces
of the color-singlet quark-antiquark bilocal operator arising from integrating 
the gluon-field is taken in the derivation. With the latter approximation, 
the formula for the $p^2$-order coefficient $F_0^2$ in Ref.\cite{Simic}
is expressed in terms of an imposed ultraviolet cut-off, and the
formula can hardly be related to the well-known Pagels-Stokar formula for 
$F_0^2$ \cite{PS}. Therefore, further improvement of studying the effective 
chiral Lagrangian from the fundamental principles of QCD is necessary.  
Actually, the study can be divided into two steps. The first step is to 
formally derive the effective chiral Lagrangian from the fundamental 
principles of QCD and express the coefficients in terms of certain dynamical 
quantities in QCD, which gives the QCD meanings of the coefficients. The 
second step is to calculate the related dynamical quantities in QCD to obtain 
the values of the coefficients. This paper is mainly devoted to the first step.

In this paper, we develop certain techniques with which we are able to 
formally derive the effective chiral Lagrangian for pseudoscalar moesons from 
the first principles of QCD without taking approximation, and all the 
coefficients are expressed in terms of certain Green's functions in QCD. 
Such expressions can be regarded as the fundamental QCD definitions of the 
coefficients.
As a simple example, we show that, under certain approximations, our 
$p^2$-order coefficient $F^2_0$ reduces to the well-known approximate 
formula given by Pagels and Stokar \cite{PS}.
A systematic numerical calculation of the coefficients by solving the 
related QCD Green's functions in certain approximation (the second 
step) will be presented in a separate paper \cite{WKXW}.

This paper is organized as follows. Sec. II is on the fundamental 
generating functional in QCD. We start from it and formally integrate out the 
heavy-quark and gluon fields to obtain a formal generating functional for the 
light quark fields. In Sec. III, we introduce a bilocal auxiliary field
reflecting the light meson degrees of freedom with which we can
integrate out the light quark fields. Then we develop a technique for 
extracting the degree of freedom of the desired local field $U(x)$ for the
pseudoscalar mesons from the bilocal auxiliary field, and formally
integrate out the the remaining degrees of freedom of the bilocal auxiliary 
field to obtain a generating functional for the local field $U(x)$. In Sec. 
IV, we develop  certain techniques to work out the complete $U(x)$-dependence 
of the effective Lagrangian in the sense of momentum expansion, and obtain the 
effective chiral Lagranian which is of the form given by Gasser and Leutwyler 
\cite{GL}. In this process we obtain the QCD expressions for all the 
coefficients in the effective chiral Lagrangian. A discussion on the relation 
between the present QCD definition of the $O(p^2)$ coefficients $F^2_0$ 
and the well-known Pagels-Stokar formula (an approximate result) \cite{PS} 
will be given in Sec. V.~ Sec. VI is a concluding remark.

\null\vspace{0.2cm}
\begin{center}
{\bf II. THE GENERATING FUNCTIONAL}
\end{center}

Consider a QCD-type gauge theory with $SU(N_{c})$ local gauge symmetry. Let 
$A_{\mu}^i~ (i=1,2,\cdots,N_c^2-1)$ be the gauge field, 
${\psi}_{\alpha}^{a\eta}$ and $\Psi_{\alpha}^{\bar{a}\eta}$ be , respectively, 
light and heavy fermion fields with color index ${\alpha}~~({\alpha}=1,2,
\cdots,N_{c})$, Lorentz spinor index $\eta$, light flavor index $a~~(a=1,2,
\cdots,N_f)$ and heavy flavor index $\bar{a}~~(\bar{a}=1,2,\cdots,N_{f}')$. 
For convenience, we simply call ${\psi}_{\alpha}^{a\eta}$ the "light 
quark-field", $\Psi_{\alpha}^{\bar{a}\eta}$ the "heavy quark-field" and 
$A_{\mu}^i$ the "gluon-field". Let us introduce local external sources 
$J_{\sigma\rho}$ for the composite light quark operators 
$\bar{\psi}^{\sigma}{\psi^{\rho}}$, where $\sigma$ and $\rho$ are short 
notations for the spinor and flavor indices. The external source $J$ can be 
decomposed into scalar, pseudoscalar, vector and axial-vector parts
\begin{eqnarray}                              
J(x)=-s(x) +ip(x)\gamma_5 +v\!\!\! /\;(x) +a\!\!\! /\;(x)\gamma_5 ,
\end{eqnarray}
where $s(x)$, $p(x)$, $v_{\mu}(x)$ and $a_{\mu}(x)$  are hermitian matrices,
and the light quark masses have been absorbed into the definition of
$s(x)$. The vector and axial-vector sources $v\!\!\! /\;(x)$ and
$a\!\!\! /\;(x)$ are taken to be traceless.

Since the contributions from the anomaly term to the effective chiral
Lagrangian has already been studied in Ref.\cite{Simic,Espriu}, our aim in
this paper is to study the complete normal part contributions. So, in this 
paper, we simply ignore the standard CP-violating term related to the anomaly 
by taking the $\theta$-vacuum parameter $\theta=0$.

Following Gasser and Leutwyler \cite{GL}, we start from constructing the 
following generating functional 
\begin{eqnarray}                                
Z[J]&=&\int{\cal D}\psi{\cal D}\bar{\psi}{\cal D}\Psi{\cal D}\bar{\Psi}
{\cal D}A_{\mu}\nonumber\\
&&\times\exp i{\int}d^{4}x\{{\cal L}({\psi},\bar{\psi},\Psi,
\bar{\Psi},A_{\mu})+\bar{\psi}J\psi\}\nonumber\\
 &=&\int{\cal D}\psi{\cal D}\bar{\psi}\exp\bigg\{i\int d^4 x
\bar{\psi}(i\partial\!\!\!/ +J)\psi\bigg\}\nonumber\\
&&\times\int{\cal D}\Psi{\cal D}\bar{\Psi}{\cal D}A_{\mu}
{\Delta}_{F}(A_{\mu})\nonumber\\
&&\times\exp\bigg\{ i{\int}d^{4}x\bigg[{\cal L}_{QCD}(A)
-\frac{1}{2\xi}[F^i(A_{\mu})]^{2}\nonumber\\
&&-g{\cal I}_i^{\mu}A^i_{\mu}
+\bar{\Psi}(i\partial\!\!\!/-M-gA\!\!\!\!/)\Psi\bigg]\bigg\},\label{QCDZ}
\end{eqnarray}                                
where ${\cal L}_{QCD}(A)=-\frac{1}{4}A^i_{\mu\nu}A^{i\mu\nu}$ is the gluon 
kinetic energy term, $M$ is the heavy quark mass matrix, ${\cal I}_i^{\mu}
\equiv\bar{\psi}\frac{\lambda_i}{2}{\gamma}^{\mu}{\psi}$ are colored currents 
composed of light quark fields, $-\frac{1}{2\xi}[F^i(A_{\mu})]^{2}$ is
the gauge-fixing term and ${\Delta}_{F}(A_{\mu})$ is the Fadeev-Popov 
determinant.

Let us first consider the integration over ${\cal D}\Psi{\cal D}\bar{\Psi}
{\cal D}A_{\mu}$ for a given configuration of $\psi$ and $\bar{\psi}$, i.e.
the current ${\cal I}_i^\mu$ serves as an external source in the integration 
over ${\cal D}\Psi{\cal D}\bar{\Psi}{\cal D}A_{\mu}$.
The result of such an integration can be formally written as
\begin{eqnarray}                                    
 &&\int{\cal D}\Psi{\cal D}\bar{\Psi}{\cal D}A_{\mu}
{\Delta}_{F}(A_{\mu})
\exp\bigg\{i{\int}d^{4}x\bigg[{\cal L}_{QCD}(A)\nonumber\\
&&-\frac{1}{2\xi}[F^i(A_{\mu})]^{2}-g{\cal I}_i^{\mu}A^i_{\mu}
+\bar{\Psi}(i\partial\!\!\!/-M-gA\!\!\!\!/)\Psi\bigg]\bigg\}\nonumber\\
 &&=\exp\;i\sum^{\infty}_{n=2}{\int}d^{4}x_1\cdots{d^{4}x_n}
\frac{(-i)^{n}g^{n}}{n!}\nonumber\\
&&\hspace{0.5cm}\times G_{\mu_1\cdots\mu_n}^{i_1
\cdots i_n}(x_1,\cdots,x_n){\cal I}^{\mu_1}_{i_1}
(x_1)\cdots{{\cal I}^{\mu_{n}}_{i_n}(x_n)},
\end{eqnarray}                                        
where $G_{\mu_1\cdots\mu_n}^{i_1\cdots{i_n}}$ is the
full n-point Green's function of the $A_{\mu}^i$-field containing internal
heavy quark lines and with given sources ${\cal I}_i^\mu$.
For simplicity, the gluon field integration in this paper is limited to
the topologically trivial sector. Inclusion of topologically non-trivial 
sectors only changes the intermediate results but not the final result 
(\ref{Sefffinal}).
 
By Fierz reordering, we can further diagonalize the color indices of the 
light-quark operators, and get
\begin{eqnarray}                             
 &&G_{\mu_1\cdots\mu_n}^{i_1\cdots{i_n}}
 (x_1,\cdots,x_n)[\bar{\psi}^{a_1}_{{\alpha}_1}(x_1)
 (\frac{\lambda_{i_1}}{2})_{\alpha_1\beta_1}\gamma^{\mu_1}
 {\psi}^{a_1}_{{\beta}_1}(x_1)]\cdots\nonumber\\
 &&\times[\bar{\psi}^{a_n}_{{\alpha}_n}(x_n)
(\frac{\lambda_{i_n}}{2})_{\alpha_n\beta_n}
\gamma^{\mu_n}{\psi}^{a_n}_{\beta_n}(x_n)]\nonumber\\
&&={\int}d^{4}x'_1\cdots{d^{4}x'_n}
g^{n-2}\bar{G}^{\sigma_1\cdots\sigma_n}_{\rho_1
\cdots\rho_n}(x_1,x'_1,\cdots,x_n,x'_n)\nonumber\\
&&\hspace{0.5cm}\times\bar{\psi}^{\sigma_1}_{\alpha_1}(x_1){\psi}^{\rho_1}
_{\alpha_1}(x'_1)\cdots\bar{\psi}^{\sigma_n}
 _{\alpha_n}(x_n){\psi}^{\rho_n}_{\alpha_n}(x'_n),
\end{eqnarray}                                    
where $\bar{G}^{\sigma_1\cdots\sigma_n}_{\rho_1
\cdots\rho_n}(x_1,x'_1,\cdots,x_n,x'_n)$
is a generalized Green's function containing 2n space-time points. Then
(\ref{QCDZ}) can be written as
\begin{eqnarray}                         
Z[J]&=&\int{\cal D}\psi{\cal D}\bar{\psi}\exp i\bigg\{\int d^{4}x
\bar{\psi}(i\partial\!\!\!/+J){\psi}+\label{faction}\nonumber\\
&&\sum^{\infty}_{n=2}{\int}d^{4}x_1\cdots{d^4}x_{n}
d^{4}x_{1}'\cdots{d^4}x_{n}'\frac{(-i)^n (g^2)^{n-1}}{n!}\nonumber\\
&&\times\bar{G}^{\sigma_1\cdots\sigma_n}_{\rho_1
\cdots\rho_n}(x_1,x'_1,\cdots,x_n,x'_n)
\nonumber\\
&&\times\bar{\psi}^{\sigma_1}_{\alpha_1}(x_1)
{\psi}^{\rho_1} _{\alpha_1}(x'_1)\cdots
\bar{\psi}^{\sigma_n}_{\alpha_n}(x_n){\psi}
^{\rho_n}_{\alpha_n}(x'_n)\bigg\}.\label{QCDZ'}
\end{eqnarray}                                     

\begin{center}
{\bf III. THE AUXILIARY FIELDS}
\end{center}
\null\noindent
{\bf 1. The bilocal Auxiliary Field}\\

For integrating out the light quark fields $\psi$ and $\bar{\psi}$, we 
introduce a bilocal auxiliary field $\Phi^{(a\eta)(b\zeta)}(x,x')$ by
inserting into (\ref{QCDZ'}) the following constant
\begin{eqnarray}                            
\int {\cal D}\Phi~\delta \bigg(N_c\Phi^{(a\eta)(b\zeta)}(x,x')-
\bar{\psi}^{a\eta}_\alpha(x)\psi^{b\zeta}_\alpha(x')\bigg).\label{Phi}
\end{eqnarray}
We see from (6) that the bilocal auxiliary field 
$\Phi^{(a\eta)(b\zeta)}(x,x')$ embodies the bilocal composite operator 
$\bar{\psi}^{a\eta}_\alpha(x)\psi^{b\zeta}_\alpha(x')$ which reflects
the meson fields. Inserting (\ref{Phi}) into (\ref{QCDZ'}) we get
\begin{eqnarray}                             
&&Z[J]\nonumber\\
&&=\int{\cal D}\psi{\cal D}\bar{\psi}{\cal D}\Phi\nonumber\\
&&\hspace{0.5cm}\times
\delta\bigg(N_c \Phi^{(a\eta)(b\zeta)}(x,x')-\bar{\psi}^{a\eta}_{\alpha}(x)
\psi^{b\zeta}_{\alpha}(x')\bigg)\nonumber\\
&&\hspace{0.5cm}\times\exp i\bigg\{{\int}d^{4}x
\bar{\psi}[i\partial\!\!\!/\;+J]{\psi}\nonumber\\
&&\hspace{0.5cm}+N_c\sum^{\infty}_{n=2}{\int}d^{4}x_{1}\cdots{d^4}x_{n}
d^{4}x_{1}'\cdots{d^4}x_{n}'\nonumber\\
&&\hspace{0.5cm}\times\frac{(-i)^n(N_c g^2)^{n-1}}{n!}
\bar{G}^{\sigma_1\cdots\sigma_n}_{\rho_1\cdots\rho_n}
(x_1,x'_1,\cdots,x_n,x'_n)\nonumber\\
&&\hspace{0.5cm}\times \Phi^{\sigma_1\rho_1}(x_1 ,x'_1)\cdots
\Phi^{\sigma_n\rho_n}(x_n ,x'_n)
\bigg\} .\label{defPhi}
\end{eqnarray}
The $\delta$-function in (\ref{defPhi}) can be further expressed in the 
Fourier representation
\begin{eqnarray}
&&\delta \bigg(N_c\Phi(x,x')-\bar{\psi}(x)\psi(x')\bigg)\nonumber\\
&&\sim\int{\cal D}\Pi e^{i\int d^4xd^4x'\Pi(x,x')\cdot
\big(N_c\Phi(x,x')-\bar{\psi}(x)\psi(x')\big)}.
\nonumber
\end{eqnarray}
With this we can integrate out the $\psi$ and $\bar{\psi}$ fields and get
\begin{eqnarray}                            
&&Z[J]\nonumber\\
&&=\int{\cal D}\Phi{\cal D}\Pi
\exp i\bigg\{ -iN_c{\rm Tr}\ln[i\partial\!\!\!/+J-\Pi]\nonumber\\
&&\hspace{0.5cm}
+\int d^{4}xd^{4}x'N_c \Phi^{\sigma\rho}(x,x')\Pi^{\sigma\rho}(x,x')\nonumber\\
&&\hspace{0.5cm}+N_c \sum^{\infty}_{n=2}{\int}d^{4}x_1\cdots{d^4}x_{n}
d^{4}x_{1}'\cdots{d^4}x_{n}'\nonumber\\
&&\hspace{0.5cm}\times\frac{(-i)^{n}(N_c g^2)^{n-1}}{n!}
\bar{G}^{\sigma_1\cdots\sigma_n}_{\rho_1
\cdots\rho_n}(x_1,x'_1,\cdots,x_n,x'_n)\nonumber\\
&&\hspace{0.5cm}\times\Phi^{\sigma_1\rho_1}(x_1 ,x'_1)\cdots
\Phi^{\sigma_n\rho_n}(x_n ,x'_n)\bigg\}\label{W00},
\end{eqnarray}         
where ${\rm Tr}$ is the functional trace with respect to the space-time, 
spinor and flavor indices.

 Let us define the classical field $\Pi_c$
\begin{eqnarray}                            
\Pi_c&\equiv &\frac{\int{\cal D}\Pi\; \Pi e^{iS}}{\int{\cal D}
\Pi\; e^{iS}},
\label{Xic}
\end{eqnarray} 
where $S$ is the argument on the exponential in (\ref{W00}).
Let $\Gamma_0[J,\Phi,\Pi_c]$ be the effective action for $\Pi_c$ with 
given $J$ and $\Phi$. $\Pi_c$ satisfies\footnote{In the conventional approach,
one usually introduces an external source ${\cal J}$ coupling to the field 
$\Pi$. With this, the right-hand-side of (\ref{eqPic}) euqals to $-{\cal J}$ 
\cite{Jackiw}. Eq.(\ref{eqPic}) corresponds to taking ${\cal J}=0$. Similarly, 
when taking ${\cal J}=0$, the effective action $\Gamma_0[J,\Phi,\Pi_c]$ euqals 
to the generating functional $W_0[J,\Phi,{\cal J}]|_{{\cal J}=0}$ for the 
connected Green's fuctions. This leads to the left-hand-side of 
(\ref{Gamma0def}).}

\begin{equation}                      
\frac{\partial \Gamma_0[J,\Phi,\Pi_c]}
{\partial \Pi_{c}^{\sigma\rho}(x,x')}=0\label{eqPic}.
\end{equation}
Then $\Gamma_0[J,\Phi,\Pi_c]$ is explicitly
\begin{eqnarray}                            
&&e^{i\Gamma_0[J,\Phi,\Pi_c]}\nonumber\\
&&\equiv\int{\cal D}\Pi\exp iN_c\bigg\{
-i{\rm Tr}\ln[i\partial\!\!\!/+J-\Pi]\nonumber\\
&&\hspace{0.5cm}+\int d^{4}xd^{4}x'
\Phi^{\sigma\rho}(x,x')\Pi^{\sigma\rho}(x,x')\nonumber\\
&&\hspace{0.5cm}+\sum^{\infty}_{n=2}{\int}d^{4}x_1\cdots{d^4}x_{n}
d^{4}x_{1}'\cdots{d^4}x_{n}'\nonumber\\
&&\hspace{0.5cm}\times\frac{(-i)^{n}(N_c g^2)^{n-1}}{n!}
\bar{G}^{\sigma_1\cdots\sigma_n}_{\rho_1
\cdots\rho_n}(x_1,x'_1,\cdots,x_n,x'_n)\nonumber\\
&&\hspace{0.5cm}\times 
\Phi^{\sigma_1\rho_1}(x_1 ,x'_1)\cdots \Phi^{\sigma_n\rho_n}(x_n ,x'_n)
\bigg\}.\label{Gamma0def}
\end{eqnarray}
With these symbols, we can formally carry out the integration over the 
$\Pi$-field in (\ref{W00}), and express the result by
\begin{equation}                          
Z[J]=\int{\cal D}\Phi\;\; \exp\{i\Gamma_0[J,\Phi,\Pi_c]\},
\label{W2}
\end{equation}
\\\\
\null\noindent
{\bf 2. Localization}\\

Since we are aiming at deriving the low energy effective chiral
Lagrangian in which the light mesons are approximately described by local 
fields, we need to {\it consistently} extract the local field degree of 
freedom from the bilocal auxiliary field $\Phi^{(a\eta)(b\zeta)}(x,x')$. The 
extraction should be {\it consistent} in the sense that the complete degree of 
freedom of the mesons resides in the local fields without leaving any in the 
coefficients in the chiral Lagrangian. Otherwise, it will affect the validity 
of the momentum expansion \cite{GL}. In this paper, we propose the following 
way of extraction, and we shall see in Sec. IV that it is really 
{\it consistent}.

The auxiliary field $\Phi$ introduced in (\ref{Phi}) has such a property which 
allows us to define the fields $\sigma$ and $\Omega^\prime$ related to
the scalar and pseudoscalar sectors of $\Phi$ as
\begin{eqnarray}                            
&&(\Omega'\sigma\Omega'+
\Omega^{\prime\dagger}\sigma\Omega^{\prime\dagger})^{ab}(x)
=(1)_{\zeta\eta} \Phi^{(b\zeta)(a\eta)}(x,x)\nonumber\\
&&(\Omega^\prime\sigma\Omega^\prime
-\Omega^{\prime\dagger}\sigma\Omega^{\prime\dagger})^{ab}(x)
=(\gamma_5)_{\zeta\eta}\Phi^{(b\zeta)(a\eta)}(x,x)\label{Udef0}.
\end{eqnarray}
Here the $\sigma$-field repesented by a hermitian matrix describes the 
modular degree of freedom, and the $\Omega'$-field represented by an
unitary matrix describes the phase degree of freedom, i.e.
\begin{equation}                             
\sigma^{\dagger}(x)=\sigma(x)\label{Sigmadef},\hspace{1cm}
\Omega^{\prime\dagger}(x)\Omega'(x)=1.\label{sigma}
\end{equation}
As usual, we can define $U^\prime(x)\equiv\Omega^{\prime 2}(x)$ which contains 
a $U(1)$ factor such that det$U^\prime(x)=e^{i\vartheta(x)}$, where the 
determinant is for the flavor matrix. The unitarity property of 
$U^\prime(x)$ implies that $\vartheta(x)$ is a real field. We can further 
extract out the $U(1)$ factor and define a field $U(x)$ as $U^\prime(x)\equiv 
e^{\frac{i}{N_f}\vartheta(x)}U(x)$. It is easy to see that det$U(x)=1$. 
Then we can define a new field $\Omega$ and decompose $U$ into 
\begin{equation}                             
U(x)=\Omega^2(x)\label{Omegadef},
\end{equation}
which is the conventional decomposition in the literature. This $U(x)$,
as the desired representation of $SU(N_f)_R\times SU(N_f)_L$, 
will be the nonlinear realization of the peudoscalar meson
fields in the chiral Lagrangian. 
Note that the way of introducing the $U(x)$-field is not unique but is up
to a chiral rotation which does not affect the final effective chiral 
Lagrangian since the chiral Lagrangian is chirally invariant.
The fields $\sigma$ and $\vartheta$ are intermediate fields which
will not appear in the final effective chiral Lagrangian. 
It is straightforward to subtract $\sigma$ from the two equations in 
(\ref{Udef0}) and using (\ref{Sigmadef}) to get
\begin{eqnarray}                        
&&e^{-i\frac{\vartheta(x)}{N_f}}\Omega^{\dagger}(x){\rm tr}_l[P_R \Phi^T(x,x)]
\Omega^{\dagger}(x)\nonumber\\
&&=e^{i\frac{\vartheta(x)}{N_f}}\Omega(x){\rm tr}_l[P_L 
\Phi^T(x,x)]\Omega(x),\label{Shermitian2}
\end{eqnarray}
where $P_R$ and $P_L$ are, resprectively, the projection operators onto the 
right-handed and left-handed states, the superscript $T$ stands for the
functional transposition (transposition of all indices including the 
space-time coordinates), and we have expressed the result in terms of $\Omega$.
 Eq.(\ref{Shermitian2}) builds up the relation between $\Phi(x,x)$ and $U(x)$ 
[or $\Omega(x)$]. Taking the determinant of (\ref{Shermitian2}) we can express
$\vartheta(x)$ in terms of $\Phi^T(x,x)$ as
\begin{eqnarray}                        
e^{2i\vartheta(x)}=\frac{{\rm det}\bigg[{\rm tr}_l[P_R\Phi^T(x,x)]\bigg]}
{{\rm det}\bigg[{\rm tr}_l[P_L\Phi^T(x,x)]\bigg]},\label{thetadef}
\end{eqnarray}
where tr$_l$ is the trace with respect to the spinor index. 

Eqs.(\ref{Udef0})-(\ref{Shermitian2}) describe our idea of localization. To 
realize this idea in the functional integration formalism, we need a technique 
to {\it integrate in} this information to the generating functional (\ref{W2}).
For this purpose, we start from the following functional identity for an 
operator ${\cal O}$ satisfying $\det{\cal O}=\det{\cal O}^\dagger$ 
(cf. Appendix for the proof)
\begin{eqnarray}                         
&&\int{\cal D}U\;\;\delta(U^{\dagger}U-1)\;\delta({\rm det}U-1)\nonumber\\
&&\times{\cal F}[{\cal O}]\;\delta(\Omega {\cal O}^{\dagger}\Omega
-\Omega^{\dagger}{\cal O}\Omega^{\dagger})\nonumber\\
&&=\mbox{const}\label{Uin},
\end{eqnarray}
in which $\int{\cal D}U\;\delta(U^{\dagger}U-1)\delta({\rm det}U-1)$ is an 
effective invariant integration measure and the function ${\cal F}[{\cal O}]$
is defined as
\begin{eqnarray}                        
\frac{1}{{\cal F}[{\cal O}]}\equiv{\rm det} {\cal O}~
\int{\cal D}\sigma\;
\delta({\cal O}^{\dagger}{\cal O}-\sigma^{\dagger}\sigma)
\delta(\sigma-\sigma^{\dagger}).\label{F}
\end{eqnarray}
With the special choice of
\begin{eqnarray}                         
{\cal O}(x)=e^{-i\frac{\vartheta(x)}{N_f}}{\rm tr}_l[P_R \Phi^T(x,x)],
\label{O}
\end{eqnarray}
which satisfies $\det{\cal O}=\det{\cal O}^\dagger$, 
eq.(\ref{Uin}) serves as the functional expression reflecting the relation 
(\ref{Shermitian2}). Inserting (\ref{Uin}) and (\ref{O}) into the functional 
(\ref{W2}) and taking the Fourier representation of the $\delta$-function
\begin{eqnarray}                         
&&\delta\bigg(\Omega{\cal O}^\dagger\Omega-\Omega^\dagger{\cal O}
\Omega^\dagger\bigg)\nonumber\\
&&\sim \int {\cal D}\Xi e^{-iN_c\int dx~
\Xi\cdot\big(\Omega{\cal O}^\dagger\Omega
-\Omega^\dagger{\cal O}\Omega^\dagger\big)},\label{Xi}
\end{eqnarray}
we get
\begin{eqnarray}                        
&&Z[J]
=\int{\cal D}\Phi{\cal D}U{\cal D}\Xi\;\delta(U^{\dagger}U-1)
\delta({\rm det}U-1)\nonumber\\
&&\hspace{0.5cm}\times\exp\bigg\{i\Gamma_0[J,\Phi,\Pi_c]+i\Gamma_I[\Phi]
+iN_c\int d^4x\nonumber\\
&&\hspace{0.5cm}\times{\rm tr}_f\bigg[\Xi(x)\bigg(
e^{-i\frac{\vartheta(x)}{N_f}}\Omega^{\dagger}(x){\rm tr}_l
[P_R \Phi^T(x,x)]\Omega^{\dagger}(x)\nonumber\\
&&\hspace{0.5cm}
-e^{i\frac{\vartheta(x)}{N_f}}\Omega(x){\rm tr}_l[P_L \Phi^T(x,x)]\Omega(x)
\bigg)\bigg]\bigg\},
\label{W3}
\end{eqnarray}
with
\begin{eqnarray}                               
&&e^{-i\Gamma_I[\Phi]}\nonumber\\
&&\equiv\prod_x\frac{1}{{\cal F}[{\cal O}(x)]}\nonumber\\
&&=\prod_x\bigg[\{{\rm det}[{\rm tr}_lP_R\Phi^T(x,x)]
{\rm det}[{\rm tr}_lP_L\Phi^T(x,x)]\}^{\frac{1}{2}}\nonumber\\
&&\hspace{0.5cm}
\times\int{\cal D}\sigma\;\delta\bigg(({\rm tr}_lP_R\Phi^T)({\rm tr}_lP_L
\Phi^T)-\sigma^{\dagger}\sigma\bigg)\nonumber\\
&&\hspace{0.5cm}\times\delta(\sigma-\sigma^{\dagger})\bigg]
\label{GammaIdef}.
\end{eqnarray}
In (\ref{W3}), the information about the relation (\ref{Shermitian2})
has been {\it integrated in}. 

Next, we deal with the functional integration over the $\Phi$-field.
For this purpose, we define an effective action 
$\tilde{\Gamma}[\Omega,J,\Xi,\Phi_c,\Pi_c]$ as
\begin{eqnarray}                        
&&e^{i{\tilde\Gamma}[\Omega,J,\Xi,\Phi_c,\Pi_c]}\nonumber\\
&&=\int{\cal D}\Phi\exp
\bigg\{i\Gamma_0[J,\Phi,\Pi_c]+i\Gamma_I[\Phi]+iN_c\int d^4x\nonumber\\
&&\hspace{0.5cm}\times tr_f\bigg[\Xi(x)\bigg(
e^{-i\frac{\vartheta(x)}{N_f}}\Omega^{\dagger}(x){\rm tr}_l
[P_R \Phi^T(x,x)]\Omega^{\dagger}(x)\nonumber\\
&&\hspace{0.5cm}-e^{i\frac{\vartheta(x)}{N_f}}\Omega(x){\rm tr}_l
[P_L \Phi^T(x,x)]\Omega(x)\bigg)\bigg]\bigg\},\label{Gammatdef}
\end{eqnarray}
in which the classical field $\Phi_c(x,x')$ is defined as 
\begin{eqnarray}                         
\Phi_c&=&\frac{\int{\cal D}\Phi\; \Phi e^{\tilde{S}}}{\int{\cal D}\Phi\;
e^{\tilde{S}}}\label{Sc},
\end{eqnarray}
where $\tilde{S}$ stands for the argument in the exponential in 
(\ref{Gammatdef}). $\Phi_c$ satisfies
\begin{equation}                         
\frac{\partial {\tilde\Gamma}[\Omega,J,\Xi,\Phi_c,\Pi_c]}
{\partial \Phi_{c}^{\sigma\rho}(x,x')}=0\label{eqPhic}.
\end{equation}
With these symbols, we can formally carry out the $\int{\cal D}\Phi$
integration in (\ref{W3}) and obtain
\begin{eqnarray}                         
Z[J]&=&\int{\cal D}U{\cal D}\Xi\;\delta(U^{\dagger}U-1)\delta({\rm det}U-1)
\nonumber\\
&&\times\exp\{i{\tilde\Gamma}[\Omega,J,\Xi,\Phi_c,\Pi_c]\}.
\label{W4}
\end{eqnarray}
Here we have formally {\it integrated out all the degrees of freedom in
$\Phi(x,x')$ besides the extracted local degree of freedom} $U(x)$. This
localization is different from those in the literature \cite{Cahill}.

Similar to the above procedures, we can formally integrate out the
$\Xi$-field by introducing an effective action
$S_{eff}[U,J,\Xi_c,\Phi_c,\Pi_c]$ as follows
\begin{equation}                         
e^{iS_{eff}[U,J,\Xi_c,\Phi_c,\Pi_c]}
=\int{\cal D}\Xi \exp i\tilde{\Gamma}[\Omega,J,\Xi,\Phi_c,\Pi_c]
\label{Seff},
\end{equation}
where the classical field $\Xi_c$ is defined as 
\begin{equation}                         
\Xi_c=\frac{\int{\cal D}\Xi\;\Xi \exp i\tilde{\Gamma}[\Omega,J,\Xi,\Phi_c,
\Pi_c]}
{\int{\cal D}\Xi\;\exp i\tilde{\Gamma}[\Omega,J,\Xi,\Phi_c,\Pi_c]}\label{Phic}
\end{equation}
and satisfies
\begin{equation}                          
\frac{\partial S_{eff}[U,J,\Xi_c,\Phi_c,\Pi_c]}
{\partial (\Xi_{c})^{ab}(x)}=0\label{eqXic}.
\end{equation}
Then the $\Xi$-integration in (\ref{W4}) can be formally carried out,
and we obtain
\begin{eqnarray}                          
Z[J]&=&\int{\cal D}U\;\delta(U^{\dagger}U-1)\delta({\rm det}U-1)\nonumber\\
&&\times\exp\{iS_{eff}[U,J,\Xi_c,\Phi_c,\Pi_c]\}.
\label{W7}
\end{eqnarray}
We see from (\ref{W7}) that $S_{eff}[U,J,\Xi_c,\Phi_c,\Pi_c]$ is just
{\it the action for $U$ with a given J}~\footnote{Note that $\Xi_c,~\Phi_c$ 
and $\Pi_c$ are all functionals of $U$ and J through (\ref{Phic}), (\ref{Sc}) 
and (\ref{Xic}).}.

Using (\ref{Gamma0def}), (\ref{Gammatdef}), ({\ref{eqXic}),
(\ref{eqPhic}), (\ref{Sc}) and (\ref{eqPic}), one 
can further show the following important relation
\begin{eqnarray}                           
&&\frac{d S_{eff}[U,J,\Xi_c,\Phi_c,\Pi_c]}{d J^{\sigma\rho}(x)}
\bigg|_{U \mbox{ fix}}\nonumber\\
&&=N_c\frac{\int{\cal D}\Xi\; \Phi_c^{\sigma\rho}(x,x)\;
 \exp i\tilde{\Gamma}[\Omega,J,\Xi,\Phi_c,\Pi_c]}
{\int{\cal D}\Xi\;\exp i\tilde{\Gamma}[\Omega,J,\Xi,\Phi_c,\Pi_c]}\nonumber\\
&&\equiv N_c\overline{\Phi_c^{\sigma\rho}(x,x)}\label{Wdiff}.
\end{eqnarray}
In (\ref{Wdiff}), the symbol $\overline{\Phi_c}$ denotes the functional 
average of $\Phi_c$ over the field $\Xi$ weighted by the action 
$\tilde{\Gamma}[\Omega,J,\Xi,\Phi_c,\Pi_c]$. {\it Eq.(\ref{Wdiff}) is crucial
in the derivation of the effective chiral Lagrangian.}

\null\vspace{0.1cm}
\begin{center}
{\bf IV. THE EFFECTIVE LAGRANGIAN}
\end{center}

To derive the effective chiral Lagrangian, we need to obtain the
$U(x)$- and $J$-dependence of $S_{eff}[U,J,\Xi_c,\Phi_c,\Pi_c]$. Note 
that $S_{eff}[U,J,\Xi_c,\Phi_c,\Pi_c]$ depends on $U$ and J not only 
explicitly in (\ref{Seff}) but also implicitly via $\Xi_c,\Phi_c$ and $\Pi_c$ 
through (\ref{Phic}), (\ref{Sc}) and (\ref{Xic}). The remaining task of the 
derivation of the effective chiral Lagrangian is to work out
explicitly the complete $U$- and $J$-dependence of 
$S_{eff}[U,J,\Xi_c,\Phi_c,\Pi_c]$. The procedure is described as what follows.

First we consider a chiral rotation
\begin{eqnarray}                       
J_{\Omega}(x)
&=&[\Omega(x)P_R+\Omega^{\dagger}(x)P_L]
~[J(x)+i\partial\!\!\!\! /\;]\nonumber\\
&&\times[\Omega(x)P_R+\Omega^{\dagger}(x)P_L],\nonumber\\
\Phi_{\Omega}^T(x,y)&=&[\Omega^{\dagger}(x)P_R+\Omega(x)P_L]~\Phi^T(x,y)
\nonumber\\
&&\times[\Omega^{\dagger}(y)P_R+\Omega(y)P_L],\nonumber\\
\Pi_{\Omega}(x,y)&=&[\Omega(x)P_R+\Omega^{\dagger}(x)P_L]~\Pi(x,y)\nonumber\\
&&\times[\Omega(y)P_R+\Omega^{\dagger}(y)P_L].\label{rotation}
\end{eqnarray}
The present theory is symmetric under this transformation. Since the
$\Xi$-field is introduced in (\ref{Xi}) and the operator 
$\Omega {\cal O}^\dagger\Omega-\Omega^\dagger {\cal O}\Omega^\dagger$
is invariant under the chiral rotation, there is no need to introduce 
$\Xi_\Omega$. Furthermore, since $\det\Omega=1$,
we can easily see from (\ref{thetadef}) that 
$\vartheta_\Omega(x)=\vartheta(x)$. The explicit dependence of 
$S_{eff}[U,J,\Xi_c,\Phi_c,\Pi_c]$ on $U(x)$ comes from the explicit 
$\Omega(x)~(\Omega^\dagger(x))$-dependence of $\tilde\Gamma[\Omega,J,\Phi_c,
\Pi_c]$ in (\ref{Gammatdef}) [cf. (\ref{Gammatdef}) and (\ref{Seff})]. After
the chiral rotation, this term becomes
\begin{eqnarray}                          
&&+iN_c\int d^4x\; tr_f\bigg[\Xi(x)\bigg(
e^{-i\frac{\vartheta(x)}{N_f}}{\rm tr}_l[P_R \Phi^T_{\Omega}(x,x)]\nonumber\\
&&-e^{i\frac{\vartheta(x)}{N_f}}{\rm tr}_l[P_L \Phi^T_{\Omega}(x,x)]\bigg)
\bigg],
\nonumber
\end{eqnarray}
which no longer depends on $U(x)$ explicitly. Therefore, after the chiral 
rotation, there is no explicit $U(x)$-dependence of
$S_{eff}[U,J,\Xi_c,\Phi_c,\Pi_c]$, i.e. the complete $U(x)$-dependence resides
implicitly in the rotated variables with the subscript $\Omega$. For instance,
the effective actions $\Gamma_0[J,\Phi_c,\Pi_c]$, $\Gamma_I[\Phi]$, 
$\tilde{\Gamma}[\Omega,J,\Xi,\Phi_c,\Pi_c]$ and $S_{eff}[U,J,\Xi_c,\Phi_c,
\Pi_c]$ can be written as
\begin{eqnarray}                             
&&\Gamma_0[J,\Phi,\Pi_c]= \Gamma_0[J_{\Omega},\Phi_{\Omega},\Pi_{\Omega c}]
+\mbox{anomaly terms},\label{Gamma0rot} \\
&&\Gamma_I[\Phi]=\Gamma_I[\Phi_{\Omega}]\label{GammaTrot},\\
&&\tilde{\Gamma}[\Omega,J,\Xi,\Phi_c,\Pi_c]= 
\tilde{\Gamma}[1,J_{\Omega},\Xi,\Phi_{\Omega c},\Pi_{\Omega c}]\nonumber\\
&&\hspace{3.5cm}+\mbox{anomaly terms}\label{Gammatilderot},\\
&&S_{eff}[U,J,\Xi_c,\Phi_c,\Pi_c]=
S_{eff}[1,J_{\Omega},\Xi_c,\Phi_{\Omega c},\Pi_{\Omega c}]\nonumber\\
&&\hspace{4.5cm}+\mbox{anomaly terms}\label{Seffrot}.
\end{eqnarray}
From (\ref{Gammatdef}) we see that
\begin{eqnarray}                       
&&e^{i\tilde{\Gamma}[1,J_{\Omega},\Xi,\Phi_{\Omega c},\Pi_{\Omega c}]}
\nonumber\\
&&=\int{\cal D}\Phi_{\Omega}\exp\bigg\{
i\Gamma_0[J_{\Omega},\Phi_{\Omega},\Pi_{\Omega c}]+i\Gamma_I[\Phi_{\Omega}]
\nonumber\\
&&\hspace{0.5cm}+iN_c\int d^4x{\rm tr}_{lf}
\{\Xi(x)[-i\sin\frac{\vartheta(x)}{N_f}\nonumber\\
&&\hspace{0.5cm}+\gamma_5\cos\frac{\vartheta(x)}{N_f}]
\Phi_{\Omega}^T(x,x)\}\bigg\}\label{tildeGammadef1},
\end{eqnarray}
where tr$_{lf}$ denotes the trace with respect to the spinor and flavor 
indices. The anomaly terms in (\ref{Gamma0rot}), (\ref{Gammatilderot}) and
(\ref{Seffrot}) are all the same arising from the non-invariance of 
Tr$\ln[i\partial\!\!\!/+J-\Pi]$ under the chiral rotation.
Note that the functional integration measure does not change under the chiral 
rotation, i.e. ${\cal D}\Phi{\cal D}\Pi={\cal D}\Phi_{\Omega}{\cal D}
\Pi_{\Omega}$ since the Jaccobians from $\Phi\to \Phi_{\Omega}$ and 
$\Pi\to \Pi_{\Omega}$ cancel each other. We see that the 
$U(x)$-dependence is simplified after the chiral rotation.

The second approach is the use of eq.(\ref{Wdiff}). As we have
mentioned in Sec. II that we ignore the irrelevant anomaly terms in
this study. Then after the chiral rotation, eq.(\ref{Wdiff}) becomes
\begin{eqnarray}                           
&&\frac{d S_{eff}[1,J_{\Omega},\Xi_c,\Phi_{\Omega c},\Pi_{\Omega c}]}
{d J^{\sigma\rho}_{\Omega}(x)}
\bigg|_{U \mbox{ fix, anomaly ignored}}\nonumber\\
&&=N_c\overline{\Phi_{\Omega c}
^{\sigma\rho}(x,x)}\label{Seffdiff}.
\end{eqnarray}
We see from (\ref{Seffdiff}) that once the $J_{\Omega}$-dependence of 
$\overline{\Phi_{\Omega c}}$ is explicitly known, one can integrate
(\ref{Seffdiff}) over $J_\Omega$ and get the $U(x)$-dependence of
$S_{eff}[1,J_\Omega,\Xi_c,\Phi_{\Omega c},\Pi_{\Omega c}]$ up to an irrelevant 
integration constant independent of $U(x)$ and $J(x)$. From that we
can derive the effective chiral Lagrangian and the expressions for its 
coefficients. There can be two ways of figuring out the $J_\Omega$-dependence 
of $\overline{\Phi_{\Omega c}}$. One is to write down the dynamical equations 
for the intermediate fields $\Xi_c,~\Phi_{\Omega c}$ and $\Pi_{\Omega c}$ and 
solve them (usually this can be done only under certain approximations)
to get the $J_\Omega$-dependence of these intermediate fields. The 
other one is to track back to the original QCD expression for the chirally 
rotated generating functional (\ref{QCDZ}) through (\ref{Wdiff}) and (\ref{Sc})
by reverting the procedures in Sec. III and Sec. II, which can lead to the 
fundamental QCD definitions of the chiral Lagrangian coefficients without 
taking approximation. We take the latter approach in this paper. Because of 
the $\delta$-function $\delta\bigg(N_c\Phi^{(a\eta)(b\zeta)}(x,x')
-\bar{\psi}^{(a\eta0}_\alpha(x)\psi^{(b\zeta)}_\alpha(x')\bigg)$ in 
(\ref{defPhi}), we can express 
$\overline{\Phi_{\Omega c}^{(a\eta)(b\zeta)}(x,y)}$ as 
\begin{eqnarray}                              
&&N_c\overline{\Phi_{\Omega c}^{(a\eta)(b\zeta)}(x,y)}=\label{SVcmeaning}\\
&&\frac{\int{\cal D}\psi{\cal D}\bar{\psi}{\cal D}\Psi
{\cal D}\bar{\Psi}{\cal D}A_{\mu}{\cal D}\Xi\bar{\psi}_{\alpha}^{a\eta}(x)
\psi_{\alpha}^{b\zeta}(y) 
e^{i\hat{S}[\psi,\bar{\psi},\Psi,\bar{\Psi},A,\Xi]}}
{\int{\cal D}\psi{\cal D}\bar{\psi}{\cal D}\Psi{\cal D}\bar{\Psi}
{\cal D}A_{\mu}{\cal D}\Xi\;
e^{i\hat{S}[\psi,\bar{\psi},\Psi,\bar{\Psi},A,\Xi]}}\nonumber
\end{eqnarray}
where
\begin{eqnarray}                         
&&\hat{S}[\psi,\bar{\psi},\Psi,\bar{\Psi},A,\Xi]\nonumber\\
&&\equiv \Gamma_I[\frac{1}{N_c}\bar{\psi}\psi]
+{\int}d^{4}x\{{\cal L}({\psi},\bar{\psi},\Psi,\bar{\Psi},A_{\mu})\nonumber\\
&&\hspace{0.5cm}+\bar{\psi}[v\!\!\! /\;_{\Omega}+a\!\!\! /\;_{\Omega}\gamma_5
-s_{\Omega}+ip_{\Omega}\gamma_5\nonumber\\
&&\hspace{0.5cm}+\Xi(-i\sin\frac{\vartheta'}{N_f}
+\gamma_5\cos\frac{\vartheta'}{N_f})]\psi\}.\label{Ihatdef}
\end{eqnarray}
{\it In (\ref{SVcmeaning}) and all later equations in this paper, the symbol
$\psi$ is used as a short notation for the chirally rotated quark field 
$\psi_\Omega$}\footnote{As an integration variable, with or without the
subscript $\Omega$ makes no difference. Once the classical equation of
motion is concerned, distinguishing the rotated $\psi_\Omega$ from the 
unrotated $\psi$ will be necessary.}.
In (\ref{Ihatdef}), $\Gamma_I[\frac{1}{N_c}\bar{\psi}\psi]$ and $\vartheta'$ 
are the quantities defined in (\ref{GammaIdef}) and (\ref{thetadef}) 
expressed in terms of quark fields, i.e.
\begin{eqnarray}                        
&&e^{-i\Gamma_I[\frac{1}{N_c}\bar{\psi}\psi]}\nonumber\\
&&=\prod_x\bigg\{
\bigg[{\rm det}\{\frac{1}{N_c}{\rm tr}_{lc}[\psi_R(x)\bar{\psi}_L(x)]\}
\nonumber\\
&&\hspace{0.5cm}\times
{\rm det}\{\frac{1}{N_c}{\rm tr}_{lc}[\psi_L(x)\bar{\psi}_R(x)]\}
\bigg]^{\frac{1}{2}}\nonumber\\
&&\hspace{0.5cm}\times\int{\cal D}\sigma\;\delta\bigg(
\frac{1}{N_c^2}{\rm tr}_{lc}(\psi_R\bar{\psi}_L)
{\rm tr}_{lc}(\psi_L\bar{\psi}_R)
-\sigma^{\dagger}\sigma\bigg)\nonumber\\
&&\hspace{0.5cm}\times\delta(\sigma-\sigma^{\dagger})\bigg\}
\label{GammaTK}
\end{eqnarray}
\begin{eqnarray}
e^{2i\vartheta'(x)}&\equiv&\frac{{\rm det}
\bigg[{\rm tr}_{lc}[\psi_R(x)\bar{\psi}_L(x)]\bigg]}
{{\rm det}\bigg[{\rm tr}_{lc}[\psi_L(x)\bar{\psi}_R(x)]\bigg]}\label{thetaKp},
\end{eqnarray}
where tr$_{lc}$ is the trace with respect to the spinor and color indices.  
(\ref{thetaKp}) implies that the range of $\vartheta'(x)$ is $[0,\pi)$.

Note that instantons contribute to both (\ref{GammaTK}) and (\ref{thetaKp}) 
\cite{tHooft}. The $U_A(1)$ violating field-configurations only cause
nonvanishing $\vartheta'$ but do not contribute to 
$\Gamma_I[\frac{1}{N_c}\bar{\psi}\psi]$. 

With (\ref{SVcmeaning})-(\ref{thetaKp}), one can integrate
(\ref{Seffdiff}) over the rotated souces and obtain
\begin{eqnarray}                                
&&e^{iS_{eff}[1,J_{\Omega},\Xi_c,\Phi_{\Omega c},\Pi_{\Omega c}]}\bigg|_{
\mbox{anomaly ignored}}\nonumber\\
&&=\int{\cal D}\psi{\cal D}\bar{\psi}{\cal D}\Psi{\cal D}\bar{\Psi}
{\cal D}A_{\mu}{\cal D}\Xi\;\; \exp\bigg\{
i\Gamma_I[\frac{1}{N_c}\overline{\psi}\psi]\nonumber\\
&&\hspace{0.5cm}+i{\int}d^{4}x
\{{\cal L}({\psi},\bar{\psi},\Psi,\bar{\Psi},A_{\mu})\nonumber\\
&&\hspace{0.5cm}+\overline{\psi}[v\!\!\! /\;_{\Omega}
+a\!\!\! /\;_{\Omega}\gamma_5 -s_{\Omega}+ip_{\Omega}\gamma_5\nonumber\\
&&\hspace{0.5cm}+\Xi(-i\sin\frac{\vartheta'}{N_f}
+\gamma_5\cos\frac{\vartheta'}{N_f})]\psi\}\bigg\}\label{Sefffin}.
\end{eqnarray}
For realistic QCD ($N_f=3$), $\cos(\vartheta'/N_f)$ does not vanish. We can 
then shift the integration variable 
$\Xi\rightarrow\Xi-ip_{\Omega}/\cos(\vartheta'/N_f)$ to cancel the 
$p_{\Omega}$-dependence in the pseudoscalar part of (\ref{Sefffin}).
After carrying out the integration over $\Xi$, we obtain
\begin{eqnarray}                                     
&&e^{iS_{eff}[1,J_{\Omega},\Xi_c,\Phi_{\Omega c},\Pi_{\Omega c}]}\bigg|_{
\mbox{anomaly ignored}}\nonumber\\
&&=\int{\cal D}\psi{\cal D}\bar{\psi}{\cal D}\Psi{\cal D}\bar{\Psi}
{\cal D}A_{\mu}\nonumber\\
&&\hspace{0.5cm}\times
\delta\bigg(\bar{\psi}^a\big(-i\sin\frac{\vartheta'}{N_f}
+\gamma_5\cos\frac{\vartheta'}{N_f}\big)\psi^b\bigg)
\nonumber\\
&&\hspace{0.5cm}\times\exp\bigg\{
i\Gamma_I[\frac{1}{N_c}\overline{\psi}\psi]+i{\int}d^{4}x
\{{\cal L}({\psi},\bar{\psi},\Psi,\bar{\Psi},A_{\mu})\nonumber\\
&&\hspace{0.5cm}+\bar{\psi}[v\!\!\! /\;_{\Omega}
+a\!\!\! /\;_{\Omega}\gamma_5 -s_{\Omega}
-p_{\Omega}\tan\frac{\vartheta'}{N_f}]\psi\}\bigg\}\label{Sefffinal}.
\end{eqnarray}
In (\ref{Sefffinal}), there is no $p_{\Omega}$-dependence in the pseudoscalar 
channel, and the $p_\Omega$-dependence appears in the scalar channel as the 
combination $s_{\Omega}+p_{\Omega}\tan\frac{\vartheta'}{N_f}$. 

Eq.(\ref{Sefffinal}) shows that {\it $S_{eff}[1,J_\Omega,,\Xi_c,\Phi_{\Omega
c},\Pi_{\Omega c}]$ is the QCD generating functional for the rotated 
sources $s_{\Omega}+p_{\Omega}\tan\frac{\vartheta'}{N_f}$,
$v^\mu_\Omega$ and $a^\mu_\Omega$ with a special parity odd degree of 
freedoms $-i\overline{\psi}^a\psi^b\sin\frac{\vartheta'}{N_f}$
$+\overline{\psi}^a\gamma_5\psi^b\cos\frac{\vartheta'}{N_f}$ frozen}. After 
making a further $U_A(1)$ rotation of the $\psi$ and $\bar{\psi}$ fields, the 
angle $\frac{\vartheta'}{N_f}$ can be rotated away and {\it the frozen degree 
of freedom becomes just the pesudoscalar degree of freedom 
$\bar{\psi}^a\gamma_5\psi^b$ as it should be}
since this degree of freedom is already included in the integrating in
of the $U$-field. The automatic occurance of this frozen degree of freedom
in the present approach implies that our way of extracting the
$U$-field degree of freedom is really {\it consistent}, i.e. nothing of the
pseudoscalar degree of freedom is left outside $U$. After the $U_A(1)$ 
rotation, $\Gamma_I$ and the Jacobian due to the rotation will give rise to an 
extra factor in the integrand, which is the compensation factor for the 
extraction of the $U$-field degree of freedom. From the point of view of the  
auxiliary field $\Phi$, this corresponds to the contributions
from integrating out the degrees of freedom other than $U$, say the 
$\sigma$ and $\eta'$ mesons.

Now we are ready to explicitly work out the effective chiral Lagrangian
to the $p^2$- and $p^4$-order. As is pointed out in Ref.\cite{GL}, the
vector and axial-vector sources should be regarded as $O(p)$ and the
scalar and pseudoscalar sources should be regarded as $O(p^2)$ in the
momentum expansion.
\\\\
\null\noindent
{\bf 1. The $p^2$-Terms}\\

We first consider the $p^2$-order terms.  To this order, the anomaly can be 
safely ignored. Expanding (\ref{Sefffinal}) up to the order of $p^2$,
we obtain
\begin{eqnarray}                                 
&&S_{eff}[1,J_{\Omega},\Xi_c,\Phi_{\Omega c},\Pi_{\Omega c}]
\bigg|_{p^2\mbox{-order}}\nonumber\\
&&=\int d^4x{\rm  tr}_f[F^{ab}(x)s_{\Omega}^{ab}(x)+F^{\prime ab}(x)
p_{\Omega}^{ab}(x)]\nonumber\\
&&\hspace{0.5cm}+\int d^4x d^4z G^{abcd}_{\mu\nu}(x,z)
a_{\Omega}^{\mu,ab}(x)a_{\Omega}^{\nu,cd}(z),\label{p2}
\end{eqnarray}
where
\begin{eqnarray}                               
&&F^{ab}(x)=-\bigg\langle[\overline{\psi}^a(x)\psi^b(x)]
\bigg\rangle\nonumber\\
&&F^{\prime ab}(x)=-\bigg\langle
[\overline{\psi}^a(x)\psi^b(x)]\tan\frac{\vartheta'(x)}{N_f}\bigg\rangle
\nonumber\\
&&G^{abcd}_{\mu\nu}(x,z)\nonumber\\
&&=\frac{i}{2}\bigg[\bigg\langle[\overline{\psi}^a(x)
\gamma_{\mu}\gamma_5\psi^b(x)]
[\overline{\psi}^c(z)\gamma_{\nu}\gamma_5\psi^d(z)]\bigg\rangle
\nonumber\\
&&\hspace{0.5cm}-\bigg\langle[\bar{\psi}^a(x)\gamma_{\mu}\gamma_5\psi^b(x)]
\bigg\rangle\bigg\langle[\bar{\psi}^c(z)\gamma_{\nu}
\gamma_5\psi^d(z)]\bigg\rangle\bigg],\label{FG}
\end{eqnarray}
and the symbol $\langle O\rangle$ for an operator $O$ appeared in (\ref{FG}) 
is defined as
\begin{eqnarray}                         
&&\displaystyle
\bigg\langle O\bigg\rangle\equiv
\frac{\int{\cal D}\mu~O}{\int{\cal D}\mu}
\label{averagedef},
\end{eqnarray}
where
\begin{eqnarray}
{\cal D}\mu&\equiv&{\cal D}\psi{\cal D}
\bar{\psi}{\cal D}\Psi{\cal D}\bar{\Psi}{\cal D}A_{\mu}\nonumber\\
&&\times\delta\bigg(\bar{\psi}^a\big(-i\sin\frac{\vartheta'}{N_f}
+\gamma_5\cos\frac{\vartheta'}{N_f}\big)\psi^b\bigg)\nonumber\\ 
&&\times e^{i\Gamma_I[\frac{1}{N_c}\overline{\psi}\psi]+
i\int d^{4}x{\cal L}({\psi},\bar{\psi},\Psi,\bar{\Psi},A_{\mu})}\nonumber.
\end{eqnarray}

For $F^{ab}(x)$, translational invariance and flavor conservation \cite{WV} 
leads to the conclusion that it is simply a space-time independent constant 
proportional to $\delta^{ab}$. So that it can be written as
\begin{eqnarray}                                   
F^{ab}(x)=F_0^2B_0\delta^{ab},
\label{lFab}
\end{eqnarray}
where
\begin{eqnarray}                                   
F_0^2B_0\equiv -\frac{1}{N_f}
\bigg\langle\bar{\psi}\psi\bigg\rangle
\label{F0B0def}
\end{eqnarray}

For $F^{\prime ab}(x)$, parity conservation \cite{WV0} leads to
\begin{eqnarray}                                  
F^{\prime ab}(x)=0.\label{Fpdef}
\end{eqnarray}

For $G^{abcd}_{\mu\nu}(x,z)$, translational invariance leads to the conclusion
that it can only depend on $x-z$. We can further expand this dependence in 
terms of $\delta(x-z)$ and its derivatives. To $p^2$-order, the derivative 
terms do not contribute, and the only term left is 
$\delta(x-z)\int d^4zG^{abcd}_{\mu\nu}(x,z)$. The coefficient 
$\int d^4zG^{abcd}_{\mu\nu}(x,z)$ is again independent of the space-time
coordinates due to translational invariance. Then Lorentz 
and flavor symmetries imply that $\int d^4xG^{abcd}_{\mu\nu}$ is
proportional to $g_{\mu\nu}\delta^{ad}\delta^{bc}$. There cannot be
terms of the structure $\delta^{ab}\delta^{cd}$ since this term is to be 
multiplied by $a^{\mu,ab}_\Omega a^{\nu,cd}_\Omega$, and $a^\mu_\Omega$ is 
traceless. Therefore the only relevant part of $G^{abcd}_{\mu\nu}(x,z)$ is 
\begin{eqnarray}                                 
G^{abcd}_{\mu\nu}(x,z)&=&
\delta(x-z)g_{\mu\nu}\delta^{ad}\delta^{bc}F_0^2\nonumber\\
&&+\mbox{irrelevant terms},
\label{F0def1}
\end{eqnarray}
where
\begin{eqnarray}                                
&&F_0^2
\equiv \frac{1}{4(N_f^2-1)}\int d^4x
[G^{\mu',abba}_{\mu'}(0,x)-\frac{1}{N_f}G^{\mu',aabb}_{\mu'}(0,x)]\nonumber\\
&&=\frac{i}{8(N_f^2-1)}\int d^4x\bigg[
\bigg\langle[\overline{\psi}^a(0)\gamma^{\mu}\gamma_5\psi^b(0)]
[\bar{\psi}^b(x)\gamma_{\mu}\gamma_5\psi^a(x)]\bigg\rangle
\nonumber\\
&&-\frac{1}{N_f}
\bigg\langle[\overline{\psi}^a(0)\gamma^{\mu}\gamma_5\psi^a(0)]
[\overline{\psi}^b(x)\gamma_{\mu}\gamma_5\psi^b(x)]\bigg\rangle
\nonumber\\
&&-\bigg\langle[\overline{\psi}^a(0)\gamma^{\mu}\gamma_5\psi^b(0)]
\bigg\rangle
\bigg\langle[\overline{\psi}^b(x)\gamma_{\mu}\gamma_5\psi^a(x)]
\bigg\rangle
\nonumber\\
&&+\frac{1}{N_f}
\bigg\langle[\overline{\psi}^a(0)\gamma^{\mu}\gamma_5\psi^a(0)]
\bigg\rangle
\bigg\langle[\overline{\psi}^b(x)\gamma_{\mu}\gamma_5\psi^b(x)]
\bigg\rangle\bigg]\nonumber\\
&&\label{F0def}.
\end{eqnarray}

Note that there is no term like tr$_f[v_{\Omega}^2]$ in (\ref{p2}). The
reason is that there exists a hidden symmetry 
$s_{\Omega}\rightarrow h^{\dagger}s_{\Omega}h$,~
$p_{\Omega}\rightarrow h^{\dagger}p_{\Omega}h$,~
$a_{\Omega}^{\mu}\rightarrow h^{\dagger}a_{\Omega}^{\mu}h$,~
and $v_{\Omega}\rightarrow h^{\dagger}v_{\Omega}^{\mu}h
+h^{\dagger}i\partial^{\mu}h$ in which the vector source transforms
inhomogeneously. So that 
the vector source can only appear together with the derivative 
$i\partial^{\mu}$ to form a covariant derivative, and a hidden symmetry
covariant quadratic form of the covariant derivative can only be an
antisymmetric tensor [cf. (\ref{covdef})] which does not contribute when 
multiplied by a symmetric coefficient of the type of (\ref{F0def1}).

With (\ref{F0B0def}), (\ref{Fpdef}) and (\ref{F0def1}) the effective action
(\ref{p2}) is then
\begin{eqnarray}                      
&&S_{eff}[1,J_{\Omega},\Xi_c,\Phi_{\Omega c},\Pi_{\Omega c}]
\bigg|_{p^2\mbox{-order}}\nonumber\\
&&=F_0^2\int d^4x{\rm  tr}_f[a_{\Omega}^2+B_0s_{\Omega}]
\nonumber\\ 
&&=F_0^2\int d^4x{\rm tr}_f
\bigg[\frac{1}{4}[\nabla^{\mu}U^{\dagger}][\nabla_{\mu}U]\nonumber\\
&&\hspace{0.5cm}+\frac{1}{2}B_0
[U(s-ip)+U^{\dagger}(s+ip)]\bigg],\label{CLp2}
\end{eqnarray}
where $\nabla_\mu$ is the covariant derivative related to the external
sources defined in Ref.\cite{GL}. 
The integrand is just the $p^2$-order chiral Lagrangian given by Gasser
and Leutwyler in Ref.\cite{GL}. Now the coefficients $F^2_0$ and $B_0$ are 
defined in (\ref{F0def}) and (\ref{F0B0def}) and are expressed in terms of 
certain Green's functions of the quark fields. These can be regarded as 
{\it the fundamental QCD definitions of $F^2_0$ and $B_0$}. 
\\\\
\null\noindent
{\bf 2. The $p^4$-Terms}\\

The $p^4$-order terms can be worked out along the same line. The relevant 
terms for the normal part contributions (ignoring anomaly contributions) are
\begin{eqnarray}                        
&&S_{eff}[1,J_{\Omega},\Xi_c,\Phi_{\Omega c},\Pi_{\Omega c}]
\bigg|_{p^4\mbox{-order, normal}}\nonumber\\
&&=\int d^4x tr_f\bigg[
-{\cal K}_1[d_{\mu}a_{\Omega}^{\mu}]^2\nonumber\\
&&\hspace{0.5cm}-{\cal K}_2(d^{\mu}a_{\Omega}^{\nu}-d^{\nu}a_{\Omega}^{\mu})
(d_{\mu}a_{\Omega,\nu}-d_{\nu}a_{\Omega,\mu})
\nonumber\\
&&\hspace{0.5cm}
+{\cal K}_3[a_{\Omega}^2]^2
+{\cal K}_4a_{\Omega}^{\mu}a_{\Omega}^{\nu}a_{\Omega,\mu}a_{\Omega,\nu}
+{\cal K}_5a_\Omega^2{\rm tr}_f[a_{\Omega}^2]\nonumber\\
&&\hspace{0.5cm}
+{\cal K}_6a_{\Omega}^{\mu}a_{\Omega}^{\nu}
{\rm tr}_f[a_{\Omega,\mu}a_{\Omega,\nu}]+{\cal K}_7s_{\Omega}^2
\nonumber\\
&&\hspace{0.5cm}
+{\cal K}_8s_{\Omega}{\rm tr}_f[s_{\Omega}]+{\cal K}_9p_{\Omega}^2
+{\cal K}_{10}p_{\Omega}{\rm tr}_f[p_{\Omega}]
\nonumber\\
&&\hspace{0.5cm}
+{\cal K}_{11}s_{\Omega}a_{\Omega}^2+{\cal K}_{12}s_{\Omega}{\rm tr}_f
[a_{\Omega}^2]
-{\cal K}_{13}V_{\Omega}^{\mu\nu}V_{\Omega,\mu\nu}
\nonumber\\
&&\hspace{0.5cm}
+i{\cal K}_{14}V_{\Omega}^{\mu\nu}a_{\Omega,\mu}a_{\Omega,\nu}
+{\cal K}_{15}p_{\Omega}d^{\mu}a_{\Omega,\mu}
\bigg] 
\label{p4},
\end{eqnarray}
where the covariant derivative $d_\mu$ and the antisymmetric tensor 
$V_{\Omega}^{\mu\nu}$ are defined as
\begin{eqnarray}                           
&&d^{\mu}a_{\Omega}^{\nu}\equiv \partial^{\mu}a_{\Omega}^{\nu}
-iv_{\Omega}^{\mu}a_{\Omega}^{\nu}
+ia_{\Omega}^{\nu}v_{\Omega}^{\mu}\nonumber\\
&&V_{\Omega}^{\mu\nu}=\partial^{\mu}v_{\Omega}^{\nu}
-\partial^{\nu}v_{\Omega}^{\mu}
-iv_{\Omega}^{\mu}v_{\Omega}^{\nu}+iv_{\Omega}^{\nu}v_{\Omega}^{\mu},
\label{covdef}
\end{eqnarray}
and the fifteen coefficients $~{\cal K}_1,\cdots {\cal K}_{15}$ are determined 
by the following integrations of the Green's functions
\begin{eqnarray}                            
&&\frac{i}{4}\int d^4 x\; x^{\mu'}x^{\nu'}\nonumber\\
&&\times\bigg\langle[\bar{\psi}^a(0)\gamma^{\mu}\gamma_5\psi^b(0)]
[\bar{\psi}^c(x)\gamma^{\nu}\gamma_5\psi^d(x)]\bigg\rangle_C
\nonumber\\
&&=[(\frac{1}{2}{\cal K}_1-{\cal K}_2)(g^{\mu\mu'}g^{\nu\nu'}
+g^{\mu\nu'}g^{\nu\mu'})\nonumber\\
&&\hspace{0.5cm}+2{\cal K}_2g^{\mu\nu}g^{\mu'\nu'}]\delta^{ad}\delta^{bc}
+\mbox{irrelavent terms}\nonumber
\end{eqnarray}
\begin{eqnarray}
&&-\frac{i}{24}\int d^4 xd^4 yd^4z\nonumber\\
&&\times\bigg\langle[\bar{\psi}^{a_1}(0)\gamma^{\mu_1}\gamma_5\psi^{b_1}(0)]
[\bar{\psi}^{a_2}(x)\gamma^{\mu_2}\gamma_5\psi^{b_2}(x)]\nonumber\\
&&\times[\bar{\psi}^{a_3}(y)\gamma^{\mu_3}\gamma_5\psi^{b_3}(y)]
[\bar{\psi}^{a_4}(z)\gamma^{\mu_4}\gamma_5\psi^{b_4}(z)]\bigg\rangle_C
\nonumber\\
&&=\frac{1}{6}\bigg\{\delta^{a_1b_2}\bigg[
\delta^{a_2b_3}\delta^{a_3b_4}\delta^{a_4b_1}[
\frac{1}{2}(g^{\mu_1\mu_2}g^{\mu_3\mu_4}\nonumber\\
&&\hspace{0.5cm}+g^{\mu_1\mu_4}g^{\mu_2\mu_3}){\cal K}_3
+g^{\mu_1\mu_3}g^{\mu_2\mu_4}{\cal K}_4]\nonumber\\
&&\hspace{0.5cm}+\delta^{a_2b_4}\delta^{a_4b_3}\delta^{a_3b_1}[
\frac{1}{2}(g^{\mu_1\mu_2}g^{\mu_3\mu_4}\nonumber\\
&&\hspace{0.5cm}+g^{\mu_1\mu_3}g^{\mu_2\mu_4}){\cal K}_3
+g^{\mu_1\mu_4}g^{\mu_2\mu_3}{\cal K}_4]\bigg]\nonumber\\
&&\hspace{0.5cm}+\delta^{a_1b_3}\bigg[
\delta^{a_3b_2}\delta^{a_2b_4}\delta^{a_4b_1}[
\frac{1}{2}(g^{\mu_1\mu_3}g^{\mu_2\mu_4}\nonumber\\
&&\hspace{0.5cm}+g^{\mu_1\mu_4}g^{\mu_2\mu_3}){\cal K}_3
+g^{\mu_1\mu_2}g^{\mu_3\mu_4}{\cal K}_4]\nonumber\\
&&\hspace{0.5cm}+\delta^{a_3b_4}\delta^{a_4b_2}\delta^{a_2b_1}[
\frac{1}{2}(g^{\mu_1\mu_3}g^{\mu_2\mu_4}\nonumber\\
&&\hspace{0.5cm}+g^{\mu_1\mu_2}g^{\mu_3\mu_4}){\cal K}_3
+g^{\mu_1\mu_4}g^{\mu_2\mu_3}{\cal K}_4]\bigg]\nonumber\\
&&\hspace{0.5cm}+\delta^{a_1b_4}\bigg[
\delta^{a_4b_2}\delta^{a_2b_3}\delta^{a_3b_1}[
\frac{1}{2}(g^{\mu_1\mu_4}g^{\mu_2\mu_3}\nonumber\\
&&\hspace{0.5cm}+g^{\mu_1\mu_3}g^{\mu_2\mu_4}){\cal K}_3
+g^{\mu_1\mu_2}g^{\mu_3\mu_4}{\cal K}_4]\nonumber\\
&&\hspace{0.5cm}+\delta^{a_4b_3}\delta^{a_3b_2}\delta^{a_2b_1}[
\frac{1}{2}(g^{\mu_1\mu_4}g^{\mu_2\mu_3}\nonumber\\
&&\hspace{0.5cm}+g^{\mu_1\mu_2}g^{\mu_3\mu_4}){\cal K}_3
+g^{\mu_1\mu_3}g^{\mu_2\mu_4}{\cal K}_4]\bigg]\nonumber\\
&&\hspace{0.5cm}+\delta^{a_1b_2}\delta^{a_2b_1}\delta^{a_3b_4}\delta^{a_4b_3}
[g^{\mu_1\mu_2}g^{\mu_3\mu_4}2{\cal K}_5\nonumber\\
&&\hspace{0.5cm}
+(g^{\mu_1\mu_3}g^{\mu_2\mu_4}+g^{\mu_1\mu_4}g^{\mu_2\mu_3}){\cal K}_6]
\nonumber\\
&&\hspace{0.5cm}+\delta^{a_1b_3}\delta^{a_3b_1}\delta^{a_2b_4}\delta^{a_4b_2}
[g^{\mu_1\mu_3}g^{\mu_2\mu_4}2{\cal K}_5\nonumber\\
&&\hspace{0.5cm}
+(g^{\mu_1\mu_2}g^{\mu_3\mu_4}+g^{\mu_1\mu_4}g^{\mu_2\mu_3}){\cal K}_6]
\nonumber\\
&&\hspace{0.5cm}+\delta^{a_1b_4}\delta^{a_4b_1}\delta^{a_2b_3}\delta^{a_3b_2}
[g^{\mu_1\mu_4}g^{\mu_2\mu_3}2{\cal K}_5\nonumber\\
&&\hspace{0.5cm}
+(g^{\mu_1\mu_2}g^{\mu_3\mu_4}+g^{\mu_1\mu_3}g^{\mu_2\mu_4}){\cal K}_6]\bigg\}
\nonumber\\
&&\hspace{0.5cm}
+\mbox{irrelavent terms},\nonumber
\end{eqnarray}
\begin{eqnarray}
&&\frac{i}{2}\int d^4 x\bigg\langle[\bar{\psi}^a(0)\psi^b(0)]
[\bar{\psi}^c(x)\psi^d(x)]\bigg\rangle_C\nonumber\\
&&={\cal K}_7\delta^{ad}\delta^{bc}+{\cal K}_8\delta^{ab}\delta^{cd}\nonumber
\end{eqnarray}
\begin{eqnarray}
&&\frac{i}{2}\int d^4 x\bigg\langle[\bar{\psi}^a(0)\psi^b(0)]
\tan\frac{\vartheta'(0)}{N_f}[\bar{\psi}^c(x)\psi^d(x)]
\nonumber\\
&&\times\tan\frac{\vartheta'(x)}{N_f}
\bigg\rangle_C
={\cal K}_9\delta^{ad}\delta^{bc}+{\cal K}_{10}\delta^{ab}\delta^{cd}
\nonumber
\end{eqnarray}
\begin{eqnarray}
&&\frac{1}{8}\int d^4 xd^4 y\bigg\langle[\bar{\psi}^{a_1}(0)
\psi^{b_1}(0)]\nonumber\\
&&\times[\bar{\psi}^{a_2}(x)\gamma^{\mu}\gamma_5
\psi^{b_2}(x)][\bar{\psi}^{a_3}(y)\gamma_{\mu}\gamma_5
\psi^{b_3}(y)]\bigg\rangle_C\nonumber\\
&&=\frac{1}{2}{\cal K}_{11}(\delta^{a_1b_2}\delta^{a_2b_3}
\delta^{a_3b_1}+\delta^{a_1b_3}\delta^{a_3b_2}\delta^{a_2b_1})\nonumber\\
&&\hspace{0.5cm}+{\cal K}_{12}\delta^{a_1b_1}\delta^{a_2b_3}\delta^{a_3b_2}
+\mbox{irrelavent terms},\nonumber
\end{eqnarray}
\begin{eqnarray}
{\cal K}_{13}&=&\frac{i}{576(N_f^2-1)}\int d^4 x\; \bigg[
(5g_{\mu\nu}g_{\mu'\nu'}-2g_{\mu\mu'}g_{\nu\nu'})\nonumber\\
&&\times x^{\mu'}x^{\nu'}[\bigg\langle\;
\overline{\psi}^a(0)\gamma^{\mu}\psi^b(0)\;
\overline{\psi}^b(x)\gamma^{\nu}\psi^a(x)\;\bigg\rangle_C\; \nonumber\\
&&-\frac{1}{N_f}\bigg\langle\;\overline{\psi}^a(0)\gamma^{\mu}\psi^a(0)\;
\overline{\psi}^b(x)\gamma^{\nu}\psi^b(x)\;\bigg\rangle_C]\;\bigg]\nonumber
\end{eqnarray}
\begin{eqnarray}
&&{\cal K}_{14}
=\frac{i}{36}\bigg[2g_{\mu_1\mu_4}g_{\mu_2\mu_3}T_A^{\mu_1\mu_2\mu_3\mu_4}
+2g_{\mu_1\mu_2}g_{\mu_3\mu_4}T_A^{\mu_1\mu_2\mu_3\mu_4}\nonumber\\
&&\hspace{0.5cm}-g_{\mu_1\mu_3}g_{\mu_2\mu_4}T_A^{\mu_1\mu_2\mu_3\mu_4}
-2g_{\mu_1\mu_4}g_{\mu_2\mu_3}T_B^{\mu_1\mu_2\mu_3\mu_4}\nonumber\\
&&\hspace{0.5cm}-2g_{\mu_1\mu_2}g_{\mu_3\mu_4}T_B^{\mu_1\mu_2\mu_3\mu_4}
+g_{\mu_1\mu_3}g_{\mu_2\mu_4}T_B^{\mu_1\mu_2\mu_3\mu_4}\bigg]\nonumber
\end{eqnarray}
\begin{eqnarray}
{\cal K}_{15}&=&\frac{i}{4(N_f^2-1)}\int d^4 x\; x^{\mu}\bigg[
\bigg\langle\;\overline{\psi}^a(0)\psi^b(0)
\tan\frac{\vartheta'(0)}{N_f}\;\nonumber\\
&&\times\overline{\psi}^b(x)\gamma_{\mu}\gamma_5\psi^a(x)
\;\bigg\rangle_C\;
-\frac{1}{N_f}\bigg\langle\;\overline{\psi}^a(0)\psi^a(0)
\tan\frac{\vartheta'(0)}{N_f}\;\nonumber\\
&&\times
\overline{\psi}^b(x)\gamma_{\mu}\gamma_5\psi^b(x)\;\bigg\rangle_C\bigg]
\label{coeffdef}
\end{eqnarray}
with $T_A,~T_B$ defined as
\begin{eqnarray}                               
&&-\frac{1}{2}\int d^4 xd^4 y\; 
x^{\mu_4}\bigg\langle\;
\overline{\psi}^{a_1}(0)\gamma^{\mu_1}\psi^{b_1}(0)\;\nonumber\\
&&\times\overline{\psi}^{a_2}(x)\gamma^{\mu_2}\gamma_5\psi^{b_2}(x)\;
\overline{\psi}^{a_3}(y)\gamma^{\mu_3}\gamma_5\psi^{b_3}(y)\;\bigg\rangle_C
\nonumber\\
&&=\delta^{a_1b_2}\delta^{a_2b_3}\delta^{a_3b_1}
T_A^{\mu_1\mu_2\mu_3\mu_4}
+\delta^{a_1b_3}\delta^{a_3b_2}\delta^{a_2b_1}
T_B^{\mu_1\mu_2\mu_3\mu_4}\nonumber\\
&&\hspace{0.5cm}+\mbox{irrelavent terms}.\label{TABdef}
\end{eqnarray}
In (\ref{coeffdef}) and ({\ref{TABdef}), $\langle\cdots\rangle_C$ denotes the 
connected part of $\langle\cdots\rangle$, and the irrelavent terms are those 
leading to tr$_f[a_{\Omega}^{\mu}]$ or tr$_f[v_{\Omega}^{\mu}]$ after 
multiplied by the corresponding sources. 

To further evaluate the rotated source parts in (\ref{p4}), we
make use of the $p^2$-order equation of motion
\begin{eqnarray}                                 
d_{\mu}a_{\Omega}^{\mu}=-B_0[p_{\Omega}-\frac{1}{N_f}{\rm tr}_f(p_{\Omega})]
\label{EM}
\end{eqnarray}
and the following identities
\begin{eqnarray}                                 
&&d^{\mu}a_{\Omega}^{\nu}-d^{\nu}a_{\Omega}^{\mu}
=\frac{1}{2}[\Omega^{\dagger}F_R^{\mu\nu}\Omega-\Omega F_L^{\mu\nu}
\Omega^{\dagger}]
\nonumber\\
&&a_{\Omega}^{\mu}=\frac{i}{2}\Omega^{\dagger}[\nabla^{\mu}U]\Omega^{\dagger}
(x)\nonumber\\
&&s_{\Omega}=\frac{1}{2}[\Omega(s-ip)\Omega+\Omega^{\dagger}(s+ip)
\Omega^{\dagger}]\nonumber\\
&&p_{\Omega}=\frac{i}{2}[\Omega(s-ip)\Omega
-\Omega^{\dagger}(s+ip)\Omega^{\dagger}]\nonumber\\
&&V_{\Omega}^{\mu\nu}=\frac{1}{2}[\Omega^{\dagger}F_R^{\mu\nu}\Omega+\Omega 
F_L^{\mu\nu}\Omega^{\dagger}]
\nonumber\\
&&\hspace{1.3cm}
+\frac{i}{4}\Omega^{\dagger}[-(\nabla^{\mu}U)U^{\dagger}(\nabla^{\nu}U)
\nonumber\\
&&\hspace{1.3cm}
+(\nabla^{\nu}U)U^{\dagger}(\nabla^{\mu}U)]\Omega^{\dagger},\label{ID}
\end{eqnarray}
where $F_{R}^{\mu\nu}$ and $F_L^{\mu\nu}$ are, respectively, the 
field-strength tensors of the right-handed and left-handed souces defined 
in Ref.\cite{GL}. With (\ref{EM}) and (\ref{ID}) and taking $N_f=3$,
eq. (\ref{p4}) becomes
\begin{eqnarray}                           
&&S_{eff}[1,J_\Omega,\Xi_c,\Phi_{\Omega c},\Pi_{\Omega
c}]\bigg|_{p^4-order,~{\mbox normal}}\nonumber\\
&&=\int d^4x\bigg[L^{(norm)}_1[{\rm tr}_f(\nabla^{\mu}U^{\dagger}\nabla_{\mu}U)]^2
\nonumber\\
&&\hspace{0.5cm}
+L^{(norm)}_2{\rm tr}_f[\nabla_{\mu}U^{\dagger}\nabla_{\nu}U]
{\rm tr}_f[\nabla^{\mu}U^{\dagger}\nabla^{\nu}U]\nonumber\\
&&\hspace{0.5cm}+L^{(norm)}_3{\rm tr}_f[(\nabla^{\mu}U^{\dagger}\nabla_{\mu}U)^2]
\nonumber\\
&&\hspace{0.5cm}
+L^{(norm)}_4{\rm tr}_f[\nabla^{\mu}U^{\dagger}\nabla_{\mu}U]
{\rm tr}_f[\chi^{\dagger}U+\chi U^\dagger]\nonumber\\
&&\hspace{0.5cm}
+L^{(norm)}_5{\rm tr}_f[\nabla^{\mu}U^{\dagger}\nabla_{\mu}U(\chi^{\dagger}U
+U^\dagger\chi)]
\nonumber\\
&&\hspace{0.5cm}
+L^{(norm)}_6[{\rm tr}_f(\chi^{\dagger}U+\chi U^{\dagger})]^2\nonumber\\
&&\hspace{0.5cm}+L^{(norm)}_7[{\rm tr}_f(\chi^{\dagger}U-\chi U^{\dagger})]^2
\nonumber\\
&&\hspace{0.5cm}
+L^{(norm)}_8{\rm tr}_f[\chi^{\dagger}U\chi^{\dagger}U+\chi U^\dagger\chi U^\dagger]
\nonumber\\
&&\hspace{0.5cm}
-iL^{(norm)}_9{\rm tr}_f[F_{\mu\nu}^R\nabla^{\mu}U\nabla^{\nu}U^{\dagger}
+F_{\mu\nu}^L\nabla^{\mu}U^\dagger\nabla^{\nu}U]\nonumber\\
&&\hspace{0.5cm}+L^{(norm)}_{10}{\rm tr}_f[U^{\dagger}F_{\mu\nu}^RUF^{L,\mu\nu}]
\nonumber\\
&&\hspace{0.5cm}
+H^{(norm)}_1{\rm tr}_f[F_{\mu\nu}^RF^{R,\mu\nu}+F_{\mu\nu}^LF^{L,\mu\nu}]\nonumber\\
&&\hspace{0.5cm}+H^{(norm)}_2{\rm tr}_f[\chi^{\dagger}\chi]\bigg],\label{p4GL}
\end{eqnarray}
where $\chi\equiv 2B_0(s+ip)$~~\footnote{In Ref.\cite{GL}, $\chi$ is
defined as $\chi\equiv 2B_0(s+ip)e^{\frac{i}{3}\theta}$. In this 
paper we have taken $\theta=0$.}. The integrand in (\ref{p4GL}) is just the
normal part contributions to the $p^4$-order terms in the chiral 
Lagrangian in Ref.\cite{GL}, and the coefficients are now defined by
\begin{eqnarray}                          
L^{(norm)}_1&=&\frac{1}{32}{\cal K}_4+\frac{1}{16}{\cal K}_5+\frac{1}{16}
{\cal K}_{13}-\frac{1}{32}{\cal K}_{14}\nonumber,\\
L^{(norm)}_2&=&\frac{1}{16}({\cal K}_4+{\cal K}_6)+\frac{1}{8}{\cal K}_{13}-
\frac{1}{16}{\cal K}_{14}\nonumber,\\
L^{(norm)}_3&=&\frac{1}{16}({\cal K}_3-2{\cal K}_4-6{\cal K}_{13}
+3{\cal K}_{14})\nonumber,\\
L^{(norm)}_4&=&\frac{{\cal K}_{12}}{16B_0}\nonumber,\\
L^{(norm)}_5&=&\frac{{\cal K}_{11}}{16B_0}\nonumber,\\
L^{(norm)}_6&=&\frac{{\cal K}_8}{16B_0^2}\nonumber,\\
L^{(norm)}_7&=&-\frac{{\cal K}_1}{16N_f}-\frac{{\cal K}_{10}}{16B_0^2}
-\frac{{\cal K}_{15}}{16B_0N_f}\nonumber,\\
L^{(norm)}_8&=&\frac{1}{16}[{\cal K}_1+\frac{1}{B_0^2}{\cal K}_7
-\frac{1}{B_0^2}{\cal K}_9+\frac{1}{B_0}{\cal K}_{15}]\nonumber,\\
L^{(norm)}_9&=& \frac{1}{8}(4{\cal K}_{13}-{\cal K}_{14})\nonumber,\\
L^{(norm)}_{10}&=&\frac{1}{2}({\cal K}_2-{\cal K}_{13})\nonumber,\\
H^{(norm)}_1&=&-\frac{1}{4}({\cal K}_2+{\cal K}_{13})\nonumber,\\
H^{(norm)}_2&=&\frac{1}{8}[-{\cal K}_1+\frac{1}{B_0^2}{\cal K}_7
+\frac{1}{B_0^2}{\cal K}_9-\frac{1}{B_0}{\cal K}_{15}].\label{p4C}
\end{eqnarray}
The twelve standard coefficients $L^{(norm)}_1$, $~L^{(norm)}_2$, $\cdots$,
$L^{(norm)}_{10}$, $~H^{(norm)}_1$, $~H^{(norm)}_2$ are expressed in terms of 
twelve independent $p^4$-order coefficients, 
${\cal K}_2$, ${\cal K}_{3,4}\equiv {\cal K}_3-2{\cal K}_4$, ${\cal K}_{4,5}
\equiv {\cal K}_4+2{\cal K}_5$, ${\cal K}_{4,6}\equiv {\cal K}_4+{\cal K}_6$,
${\cal K}_7$, ${\cal K}_8$, ${\cal K}_{1,9,15}\equiv {\cal K}_1-
\frac{1}{B^2_0}{\cal K}_9+\frac{1}{B_0}{\cal K}_{15}
$, ${\cal K}_{1,10,15}\equiv {\cal K}_1+\frac{N_f}{B^2_0}{\cal K}_{10}
+\frac{1}{B_0}{\cal K}_{15}$, ${\cal K}_{11}$, ${\cal K}_{12}$, 
${\cal K}_{13}$, and ${\cal K}_{14}$.

The total coefficients are then
\begin{eqnarray}                                                 
&&L_i=L^{(norm)}_i+L^{(anom)}_i,\,\,\,\,\,\,i=1,\cdots,10,\nonumber\\
&&H_j=H^{(norm)}_j+H^{(anom)}_j,\,\,\,\,\,\,j=1,2,
\label{L}
\end{eqnarray}
where $L^{(anom)}_i$ and $H^{(anom)}_j$ are the anomaly contributions
to the coefficients given in Ref.\cite{Simic,Espriu}.

So we have formally derived the $p^4$-order terms of the 
chiral Lagrangian from the fundamental principles of QCD without taking 
approximations and have expressed all the coefficients in terms of the
integrations of certain Green's functions in QCD. {\it Eqs.(\ref{coeffdef}), 
(\ref{TABdef}) and (\ref{p4C}) give the fundamental QCD definitions of the 
the twelve coefficients $L^{(norm)}_1\cdots L^{(norm)}_{10},~H^{(norm)}_1$ and 
$H^{(norm)}_2$}. The procedure can be carried on order by order in the 
momentum expansion. 

The expressions (\ref{F0B0def}), (\ref{F0def}), (\ref{coeffdef}),
(\ref{TABdef}) and (\ref{p4C}) are convenient for lattice QCD calculations
of the fifteen coefficients.

\null\vspace{0.2cm}
\begin{center}
{\bf V. ON THE COEFFICIENTS ${\bf F^2_0}$ AND ${\bf B_0}$}
\end{center}

So far, we have given the formal QCD definitions of the fourteen coefficients 
of the chiral Lagrangian up to the $p^4$-order. To get the values of
the coefficients, we need to solve the relevant Green's functions which is a
hard task, and we shall present the calculations in a separate paper 
\cite{WKWX}. 
To have an idea of how our present formulae are related to
other known approximate results, we take the $p^2$-order coefficients $F^2_0$ 
and $B_0$ [eqs. (\ref{F0B0def}) and (\ref{F0def})] as examples and make the 
following simple discussion.

As we have mentioned in Sec. IV, there can be two ways of figuring out
the explicit $J_\Omega$-dependence of $\overline{\Phi_{\Omega c}}$ for 
evaluating $S_{eff}$ from (\ref{Seffdiff}). One of them is to solve the
dynamical equations for the intermediate fields 
$\Phi_{\Omega c},~\Pi_{\Omega c}$ and $\Xi_c$, and the other is to track 
back to the original QCD generating functional without the intermediate
fields. For convenience, we took the latter way in the above derivation 
of the chiral Lagrangian. To compare our results with the known approximate 
results, we are going to take certain approximations, say the large $N_c$ 
limit, with which the calculation of the intermediate fields becomes even more 
convenient. Thus we take the former way in the following discussion.

First we take the large $N_c$ limit. It can be easily checked that, in
this limit, the functional integrations in (\ref{Gamma0def}), 
(\ref{Gammatdef}) and (\ref{Seff}) can be simply carried out by the saddle 
point approximation (taking the classical orbit in the semiclassical 
approximation). The saddle point equations (\ref{eqPic}), (\ref{eqPhic}) and
(\ref{eqXic}) are just the dynamical equations determining 
$\Pi_{\Omega c},\Phi_{\Omega c},\Xi_c$ as functions of $J_{\Omega}$,
which are
\begin{eqnarray}                           
&&\Phi^{(a\eta)(b\zeta)}_{\Omega c}(x,y)=-i[(i\partial\!\!\! /\;+J_{\Omega}
-\Pi_{\Omega c})^{-1}]^{(b\zeta)(a\eta)}(y,x),\nonumber\\
&&\tilde{\Xi}^{\sigma\rho}(x)\delta(x-y)+\Pi_{\Omega c}^{\sigma\rho}(x,y)
\nonumber\\
&&+\sum^{\infty}_{n=1}{\int}d^{4}x_1\cdots{d^4}x_{n}
d^{4}x_{1}'\cdots{d^4}x_{n}'\frac{(-i)^{n+1}(N_c g^2)^n}{n!}\nonumber\\
&&\times\overline{G}^{\sigma\sigma_1\cdots\sigma_n}_{\rho\rho_1
\cdots\rho_n}(x,y,x_1,x'_1,\cdots,x_n,x'_n)
\Phi^{\sigma_1\rho_1}_{\Omega c}(x_1 ,x'_1)\cdots\nonumber\\
&&\times\Phi^{\sigma_n\rho_n}_{\Omega c}(x_n ,x'_n)=0,\nonumber\\
&&{\rm tr}_l[(-i\sin\frac{\vartheta(x)}{N_f}
+\gamma_5\cos\frac{\vartheta(x)}{N_f})\Phi_{\Omega c}^T(x,x)]=0,
\label{saddeq}
\end{eqnarray}
where $\tilde{\Xi}$ is a short notation for the following quantity
\begin{eqnarray}                                   
\tilde{\Xi}^{\sigma\rho}(x)&\equiv&\frac{\partial}
{\partial\Phi^{\sigma\rho}_c(x,x)}
\int d^4y\; {\rm tr}_{lf}\{\Xi_c(y)[-i\sin\frac{\vartheta_c(y)}{N_f}\nonumber\\
&&+\gamma_5\cos\frac{\vartheta_c(y)}{N_f}]\Phi_{\Omega,c}^T(y,y)\}
\bigg|_{\Xi_c\mbox{ fixed}}.\label{constraint}
\end{eqnarray}
In (\ref{saddeq}) and (\ref{constraint}), $\vartheta_c(x)$ depends on 
$\Phi(x,x)$ through (\ref{thetadef}).
Note that the effective action $\Gamma_I[\Phi_{\Omega}]$ belongs to $O(1/N_c)$,
so that it does not contribute in the present approximation.
In (\ref{saddeq}), the field $\Pi_{\Omega c}$ can be easily eliminated 
and the resulting equation is
\begin{eqnarray}                            
&&[i\partial\!\!\! /\;+i\Phi_{\Omega c}^{T,-1}
+v\!\!\! /\;_{\Omega}+a\!\!\! /\;_{\Omega}\gamma_5 -s_{\Omega}
+ip_{\Omega}\gamma_5\nonumber\\
&&+\tilde{\Xi}]^{\sigma\rho}(x,y)
+\sum^{\infty}_{n=1}{\int}d^{4}x_1\cdots{d^4}x_{n}
d^{4}x_{1}'\cdots{d^4}x_{n}'\nonumber\\
&&\times\frac{(-i)^{n+1}(N_c g^2)^n}{n!}
\overline{G}^{\sigma\sigma_1\cdots\sigma_n}_{\rho\rho_1
\cdots\rho_n}(x,y,x_1,x'_1,\cdots,x_n,x'_n)\nonumber\\
&&\times\Phi_{\Omega c}^{\sigma_1\rho_1}(x_1 ,x'_1)\cdots 
\Phi_{\Omega c}^{\sigma_n\rho_n}(x_n ,x'_n)=0.\label{fineq}
\end{eqnarray}
 
In order to compare our results with the usual dynamical equations in the 
ladder approximation, we further take the ladder approximation which,
in the present case, corresponds to ignoring all the $n>1$ terms in 
(\ref{fineq}) and with
\begin{eqnarray}                            
&&\overline{G}^{\sigma_{1}\sigma_{2}}_{\rho_{1}\rho_{2}}
(x_{1},x'_{1},x_{2},x'_{2})\nonumber\\
&&=-\frac{1}{2}G_{\mu_{1}\mu_{2}}(x_{1},x_{2})
(\gamma^{\mu_{1}})_{\sigma_{1}\rho_{2}}(\gamma^{\mu_{2}})_{\sigma_{2}
\rho_{1}}\nonumber\\
&&\hspace{0.5cm}\times{\delta}(x'_{1}-x_{2}){\delta}(x'_{2}-x_{1})+
O(\frac{1}{N_{c}})\mbox{ term},\label{ladder}\nonumber
\end{eqnarray}                   
where $G_{\mu\nu}(x,y)$ is the gluon propagator without internal light-quark 
lines. 
Then, in the ladder approximation, (\ref{fineq}) becomes
\begin{eqnarray}                            
&&[i\partial\!\!\! /\;+i\Phi_{\Omega c}^{T,-1}
+v\!\!\! /\;_{\Omega}+a\!\!\! /\;_{\Omega}\gamma_5 -s_{\Omega}
+ip_{\Omega}\gamma_5\nonumber\\
&&+\tilde{\Xi}](x,y)
+\frac{1}{2}g^2N_cG_{\mu\nu}(x,y)\gamma^{\mu}\Phi_{\Omega c}^T(x,y)
\gamma^{\nu}=0\nonumber\\
&&\label{eqladdar}
\end{eqnarray}

On the other hand, in the large $N_c$ limit, $\overline{\Phi_{\Omega c}}$ is 
just $\Phi_{\Omega c}$ which is the full physical propagator of the quark with 
the rotated sources. When the sources are turned off, $\Phi_{\Omega c}$ can be 
expressed in terms of the quark self-energy $\Sigma(-p^2)$ and the wave 
function renormalization $Z(-p^2)$ by the standard expression
\begin{eqnarray}                             
&&\Phi_0^{T(a\eta)(b\zeta)}(x,y)\label{fullpropagator}\\
&&\equiv [\Phi_{\Omega c}^T]^{(a\eta)(b\zeta)}
(x,y) \bigg|_{s_{\Omega}=p_{\Omega}=v^{\mu}_{\Omega}=a^{\mu}_{\Omega}=0}
\nonumber\\
&&=\delta^{ab}\int\frac{d^4p}{(2\pi)^4}e^{-ip(x-y)}\bigg[\frac{-i}{
 Z(-p^2)p\!\!\! /\;-\Sigma(-p^2)}\bigg]^{\eta\zeta},\nonumber
\end{eqnarray}
in which translational invariance and the flavor and parity conservations
have been considered. Plugging (\ref{fullpropagator}) into (\ref{eqladdar}) 
we have
\begin{eqnarray}                             
&&-(Z(-p^2)-1)p\!\!\! /\;+\Sigma(-p^2)+\frac{1}{2}iN_cg^2
\int\frac{d^4q}{(2\pi)^4}\nonumber\\
&&\times G_{\mu\nu}(p-q)\gamma^{\mu}
\frac{1}{Z(-q^2)q\!\!\! /\;-\Sigma(-q^2)}\gamma^{\nu}=0,\label{SDeq}
\end{eqnarray}
where $G_{\mu\nu}(p)$ is the gluon propagator in the momentum representation,
and the fact $\tilde{\Xi}|_{sources=0}=0$ is taken into account. 
(\ref{SDeq}) is just the usual Schwinger-Dyson equation in the laddar 
approximation. 

With the solution of the Schwinger-Dyson equation, the formula (\ref{F0B0def}) 
for $F^2_0B_0$ can be expressed as
\begin{eqnarray}                             
F_0^2B_0=4iN_c\int\frac{d^4p}{(2\pi)^4}\frac{\Sigma(-p^2)}{Z^2(-p^2)p^2
-\Sigma^2(-p^2)}.
\end{eqnarray}
By definition [cf. (\ref{p2}) and (\ref{F0def1})], $F_0^2$ is related to the 
coefficient of the term linear in $a_{\Omega}$ in the expansion of 
$\overline{\Phi_{\Omega c}}$. 
We denote
\begin{eqnarray}                              
&&\int d^4z~\Phi_{1,\mu}^{T\eta\zeta}(\frac{x+y}{2}-z,x-y) 
a^{ab\mu}_\Omega(z)\\
&&\equiv [\Phi_{\Omega c}^T]^{(a\eta)(b\zeta)}(x,y) \bigg|_{\mbox{linear in }
a^{\mu}_{\Omega}}\nonumber\\
&&=\int d^4z\frac{d^4pd^4q}{(2\pi)^8}e^{-ip(\frac{x+y}{2}-z)-iq(x-y)}
 \Phi_{1,\mu}^{T\eta\zeta}(p,q) a^{ab\mu}_{\Omega}(z)\nonumber
 \end{eqnarray}
Then $F_0^2$ is determined by\footnote{This is an alternative
expression for $F^2_0$ equivalent to (\ref{F0def}).}
\begin{eqnarray}                            
F_0^2=\frac{N_c}{8}(\gamma^{\mu}\gamma_5)^{\eta\zeta}\int\frac{d^4q}{(2\pi)^4}
\Phi_{1,\mu}^{T\zeta\eta}(0,q).
\end{eqnarray}
To this order, we still have $\tilde{\Xi}|_{\mbox{linear in}~ a^\mu_\Omega}=0$, and eq.(\ref{ladder}) reduces to
\begin{eqnarray}                           
&&-i[\Phi_0^{T}(q+\frac{p}{2})]^{-1}\Phi^T_{1,\mu}(p,q)
[\Phi_0^T(q-\frac{p}{2})]^{-1}+\gamma_{\mu}\gamma_5\nonumber\\
&&+\frac{1}{2}g^2N_c\int\frac{d^4k}{(2\pi)^4}
G_{\mu'\nu'}(k-q)\gamma^{\mu'}\Phi^T_{1,\mu}(p,k)\gamma^{\nu'}=0.\nonumber\\
&&\label{BSeq}
\end{eqnarray}

In the literature, a further approximation of {\it dropping the last term
in (\ref{BSeq})} is usually taken \cite{solveBS} (It can be shown that
to leading order in dynamical perturbation \cite{PS}, this term can be 
reasonably ignored \cite{WKWX}.). Moreover, to leading order in dynamical
perturbation or in the Landau gauge, 
$Z(-p^2)=1$. Then $F_0^2$ becomes
\begin{eqnarray}                           
F_0^2&=&\frac{iN_c}{8}\int\frac{d^4q}{(2\pi)^4}
{\rm tr}_l[\gamma^{\mu}\gamma_5\frac{1}{q\!\!\! /\;-\Sigma(-q^2)}\gamma_{\mu}
\gamma_5\frac{1}{q\!\!\! /\;-\Sigma(-q^2)}]\nonumber\\
&=&-4iN_c\int\frac{d^4q}{(2\pi)^4}\bigg[
\frac{1}{8}\frac{\partial}{\partial q^{\mu}}
\bigg(\frac{q^{\mu}}{q^2-\Sigma^2(-q^2)}\bigg)\nonumber\\
&&+\frac{[\Sigma(-q^2)+\frac{1}{2}q^2\Sigma'(-q^2)]\Sigma(-q^2)}
{[q^2-\Sigma^2(-q^2)]^2}\bigg].\label{PSeq}
\end{eqnarray}
The anomaly contribution to $F^2_0$ calculated in Ref.\cite{Simic} is
of the same form as the first term in (\ref{PSeq}) but with an oposite
sign, so that it just cancel the first term in in (\ref{PSeq}). Then 
(\ref{PSeq}) is just the well-known Pagels-Stokar formula for $F^2_0$ \cite{PS}.
Thus the Pagels-Stokar formula is an approximate result of our formula 
by taking the approximations of the large $N_c$ limit, the ladder 
approximation and dropping the last term in (\ref{BSeq}) (or to leading
order in dynamical perturbation).

\null\vspace{0.2cm}
\begin{center}
{\bf VI. CONCLUSIONS}
\end{center}

In this paper, we have derived the normal part contributions to the 
chiral Lagrangian for pseudoscalar mesons up to the $p^4$-terms from 
the fundamental principles of QCD without taking approximations. Together with 
the anomaly part contributions given in Ref.\cite{Simic,Espriu}, it leads to 
the complete QCD theory of the chiral Lagrangian.

 We started, in Sec. II, from the fundamental generating
functional (\ref{QCDZ}) in QCD, and formally expressed the integration over the
gluon-field in terms of physical gluon Green's functions. Then we integrated 
out the quark-fields by introducing a bilocal auxiliary field $\Phi(x,y)$ [cf. 
(\ref{Phi})]. To extract the degree of freedom of the local 
pseudoscalar-meson-field $U(x)$, we developed, in Sec. III, a technique for 
{\it extracting} it from the bilocal auxiliary field $\Phi(x,y)$ [cf. 
(\ref{Udef0}) and (\ref{Omegadef})], and {\it integrating in} the extraction 
constraint into the generating functional [cf. 
(\ref{Uin}), (\ref{O}) and (\ref{W3})]. This procedure is {\it consistent} in 
the sense that the complete pseudoscalar meson degree of freedom is converted 
into the $U(x)$-field such that the pseudoscalar degree of freedom in the
quark sector is automatically frozen in the path-integral formulation
of the effective action $S_{eff}$.

 We then developed two techniques for working out the explicit 
$U(x)$-dependence of $S_{eff}$ in Sec. IV. The first one
is to introduce a chiral rotation (\ref{rotation}) which simplifies the
$U(x)$-dependence in such a way that the $U(x)$-dependence resides only
implicitly in the rotated sources and some rotated intermediate-fields, and 
the second one is to implement eq.(\ref{Seffdiff}) to obtain $S_{eff}$ from 
the averaged field $\overline{\Phi_{\Omega c}}$. To avoid dealing with
the intermediate-fields, we tracked back to the original
QCD generating functional with which the implicit $U(x)$-dependence
only resides in the rotated sources.
With all these, we expanded $S_{eff}$ in power series of the rotated
sources and explicitly derived the $p^2$-terms and $p^4$-terms
of the chiral Lagrangian for pseudoscalar mesons \cite{GL}. In this 
formulation, all the fifteen coefficients in the chiral Lagrangian are
expressed in terms of certain Green's functions in QCD [cf. (\ref{F0B0def}), 
(\ref{F0def}), (\ref{p4C}) and (\ref{coeffdef})]. {\it These formulae can be 
regarded as the fundamental QCD definitions of the fifteen coefficients in the 
chiral Lagrangian}. These expressions are convenient for lattice QCD 
calculation of the fifteen coefficients.

To see the relation between our QCD definition and the well-known
approximate results in the literature, we took the $p^2$-order
coefficients $F^2_0$ and $B_0$ as examples in Sec. V. With the
approximations of large $N_c$ limit, ladder approximation, and dropping
the momentum-integration term in the approximate Bethe-Salpeter
equation (\ref{BSeq}), our formula reduces to the well-known
Pagels-Stokar formula for the pion decay constant $F_\pi$ \cite{PS}. The 
complete calculation of the fifteen coefficients will
be presented in a separate paper \cite{WKWX}. The derivation of the effective
chiral Lagrangian further including the $\eta'$ meson or the $\rho$ meson,
and the application of the present approach to the electroweak chiral 
Lagrangian are all in preparation.

\section*{Acknowledgments}

This work is supported by the National Natural Science Foundation of China,
 and the Fundamental Research Foundation of Tsinghua University. One of
us (Q. Wang) would like to thank C.P. Yuan for the kind hospitality during his 
visit to Michigan State University.

\null\vspace{0.2cm}
\appendix
\section*{}

 Here we give the proof of eqs.(\ref{Uin}) and (\ref{F}) in the text.
 
 Consider a matrix operator ${\cal O}$ satisfying $\det{\cal
O}=\det{\cal O}^\dagger$, we calculate the following 
functional integration
\begin{eqnarray}
I&\equiv&\int{\cal D}U\;\delta(U^{\dagger}U-1)\delta({\rm det}U-1)\nonumber\\
&&\times{\cal F}[{\cal O}]~\delta(\Omega {\cal O}^{\dagger}\Omega
-\Omega^{\dagger} {\cal O}\Omega^{\dagger}),\nonumber
\end{eqnarray}
where ${\cal D}U\;\delta(U^{\dagger}U-1)\delta({\rm det}U-1)$ serves as the 
invariant integration measure at the present case. The two delta-functions 
$\delta(U^{\dagger}U-1),~\delta(\det U-1)$ constrain the 
 integration to the subspace with unitary and unity determinant of the 
$U$-field.

We can rewrite $\delta({\rm det}U-1)\delta(U^{\dagger}U-1)$ as
\begin{eqnarray}
&&\delta({\rm det}U-1)\delta(U^{\dagger}U-1)\nonumber\\
&&=\frac{1}{2}\delta([{\rm det}U]^2-1)\theta({\rm det}U)\delta(U^{\dagger}U-1) 
\nonumber\\
&&=\frac{1}{2}\delta({\rm det}U[{\rm det}U-{\rm det}U^{\dagger}])
\theta({\rm det}U)\delta(U^{\dagger}U-1)\nonumber\\
&&=\frac{1}{2}\delta({\rm det}U-{\rm det}U^{\dagger})
\frac{\theta({\rm det}U)}{{\rm det} U}\delta(U^{\dagger}U-1) 
,\nonumber
\end{eqnarray}
so that
\begin{eqnarray}                      
I&=&\frac{1}{2}\int{\cal D}U\;\delta(U^{\dagger}U-1)
\delta({\rm det}U-{\rm det}U^{\dagger})\nonumber\\
&&\times\frac{\theta({\rm det}U)}{{\rm det} U}
~{\cal F}[{\cal O}]~\delta(\Omega {\cal O}^{\dagger}\Omega
-\Omega^{\dagger} {\cal O}\Omega^{\dagger}).\nonumber
\end{eqnarray}
Next, we introduce two auxiliary fields $\Sigma$ and $\tilde{\Sigma}$
and write $I$ as
\begin{eqnarray}                       
I&=&\frac{1}{2}\int{\cal D}U{\cal D}\Sigma{\cal D}\tilde{\Sigma}\;
\delta(U^{\dagger}U-1)\delta({\rm det}U-{\rm det}U^{\dagger})\nonumber\\
&&\times\frac{\theta({\rm det}U)}{{\rm det} U}\delta(\tilde{\Sigma}-\Sigma)
~{\cal F}[{\cal O}]~\delta(\Sigma-\Omega^{\dagger} {\cal O}\Omega^{\dagger})
\nonumber\\
&&\times\delta(\tilde{\Sigma}-\Omega {\cal O}^{\dagger}\Omega)\nonumber\\
&=&\frac{1}{2}\int{\cal D}U{\cal D}[\Omega^{\dagger}\Sigma\Omega]
{\cal D}[\Omega^{\dagger}\tilde{\Sigma}\Omega]\;
\delta(U^{\dagger}U-1)\nonumber\\
&&\times\delta({\rm det}U-{\rm det}U^{\dagger})
\frac{\theta({\rm det}U)}{{\rm det} U}
~{\cal F}[{\cal O}]~\nonumber\\
&&\times\delta(\Omega^{\dagger}[\tilde{\Sigma}-\Sigma]\Omega)
\delta(\Omega^{\dagger}\Sigma\Omega-U^{\dagger}{\cal O})\nonumber\\
&&\times\delta(\Omega^{\dagger}\tilde{\Sigma}\Omega-
{\cal O}^{\dagger}U)\nonumber.
\end{eqnarray}
We then change the integration variables $\Sigma$ and $\tilde{\Sigma}$ into
\begin{eqnarray}                     
\Sigma\rightarrow\Sigma'=\Omega^{\dagger}\Sigma\Omega\hspace{0.7cm}
\tilde{\Sigma}\rightarrow\tilde{\Sigma}'
=\Omega^{\dagger}\tilde{\Sigma}\Omega,
\end{eqnarray}
and get
\begin{eqnarray}
I&=&\frac{1}{2}\int{\cal D}U{\cal D}\Sigma'{\cal D}\tilde{\Sigma}'\;
\delta(U^{\dagger}U-1)\delta({\rm det}U-{\rm det}U^{\dagger})\nonumber\\
&&\times\frac{\theta({\rm det}U)}{{\rm det}U}\delta(\tilde{\Sigma}'-\Sigma')
~{\cal F}[{\cal O}]~\delta(\Sigma'-U^{\dagger}{\cal O})\nonumber\\
&&\times\delta(\tilde{\Sigma}'-{\cal O}^{\dagger}U)\nonumber\\
&=&\frac{1}{2}\int{\cal D}U{\cal D}\Sigma'{\cal D}\tilde{\Sigma}'\;
\delta(U^{\dagger}U-1)\delta({\rm det}U-{\rm det}U^{\dagger})\nonumber\\
&&\times\frac{\theta({\rm det}U)}{[{\rm det} U]^2}\delta(\tilde{\Sigma}'
-\Sigma')~{\cal F}[{\cal O}]~\delta(U\Sigma'-{\cal O})\nonumber\\
&&\times\delta(\tilde{\Sigma}'-\Sigma^{\prime\dagger})\nonumber\\
&=&\frac{1}{2}\int{\cal D}U{\cal D}\Sigma'{\cal D}\tilde{\Sigma}'\;
{\rm det}(\Sigma^{\prime 5})
\delta(\tilde{\Sigma}'[U^{\dagger}U-1]\Sigma')\nonumber\\
&&\times\delta({\rm det}
[\tilde{\Sigma}'U]-{\rm det}[\tilde{\Sigma}'U^{\dagger}])
\frac{\theta({\rm det}U)}{[{\rm det} (U\Sigma')]^2}
\delta(\tilde{\Sigma}'-\Sigma')\nonumber\\
&&\times{\cal F}[{\cal O}]~
\delta(U\Sigma'-{\cal O})\delta(\tilde{\Sigma}'-\Sigma^{\prime\dagger})
\nonumber\\
&=&\frac{1}{2}\int{\cal D}[U\Sigma']{\cal D}\Sigma'{\cal D}\tilde{\Sigma}'\;
{\rm det}(\Sigma^{\prime 4})
\delta(\Sigma^{\prime\dagger}U^{\dagger}U\Sigma'-\Sigma^{\prime\dagger}\Sigma')
\nonumber\\
&&\times
\delta({\rm det}[U\Sigma']-{\rm det}[\Sigma^{\prime\dagger}U^{\dagger}])
\frac{\theta({\rm det}U)}{[{\rm det} (U\Sigma')]^2}\nonumber\\
&&\times\delta(\tilde{\Sigma}'-\Sigma')
~{\cal F}[{\cal O}]~\delta(U\Sigma'-{\cal O})\delta(\tilde{\Sigma}'
-\Sigma^{\prime\dagger})\nonumber.
\end{eqnarray}
We further change the integration variables $U$ into
\begin{eqnarray}                    
U\rightarrow U'=U\Sigma',
\end{eqnarray}
and get
\begin{eqnarray}
I&=&\frac{1}{2}\int{\cal D}U'{\cal D}\Sigma'{\cal D}\tilde{\Sigma}'\;
\delta(U^{\prime\dagger}U'-\Sigma^{\prime\dagger}\Sigma')\nonumber\\
&&\times
\delta({\rm det}U'-{\rm det}U^{\prime\dagger})\frac{{\rm det}
(\Sigma^{\prime 4})\theta(\frac{{\rm det}U'}{{\rm det}\Sigma'})}
{[{\rm det}(U')]^2}\nonumber\\
&&\times\delta(\tilde{\Sigma}'-\Sigma')
~{\cal F}[{\cal O}]~\delta(U'-{\cal O})
\delta(\tilde{\Sigma}'-\Sigma^{\prime\dagger}).
\nonumber
\end{eqnarray}
The $U'$ and $\tilde{\Sigma}'$ integrations can be carried out and
we obtain
\begin{eqnarray}                      
I&=&\frac{1}{2[{\rm det}{\cal O}]^2}
\delta({\rm det}{\cal O}-{\rm det}O^{\dagger})\int{\cal D}\Sigma'\;
\delta({\cal O}^{\dagger}{\cal O}-\Sigma^{\prime\dagger}\Sigma')\nonumber\\
&&\times{\rm det}(\Sigma^{\prime 4})
\theta(\frac{{\rm det}{\cal O}}{{\rm det}\Sigma'}){\cal F}[{\cal O}]
\delta(\Sigma'-\Sigma^{\prime\dagger})\nonumber\\
&=&\frac{1}{2[{\rm det}{\cal O}]^3}
\delta(1-\frac{{\rm det}{\cal O}^{\dagger}}{{\rm det}{\cal O}})
\int{\cal D}\Sigma'\;\delta({\cal O}^{\dagger}{\cal O}
-\Sigma^{\prime\dagger}\Sigma')\nonumber\\
&&\times [{\rm det}(\Sigma^{\prime\dagger}\Sigma')]^2
\theta(\frac{{\rm det}{\cal O}}{\sqrt{{\rm det}[\Sigma^{\prime\dagger}
\Sigma']}}){\cal F}[{\cal O}]\delta(\Sigma'-\Sigma^{\prime\dagger})\nonumber\\
&=&\frac{1}{2[{\rm det}{\cal O}]^2}
\delta(1-\frac{{\rm det}{\cal O}^{\dagger}}{{\rm det}{\cal O}})
\int{\cal D}\Sigma'\;
\delta({\cal O}^{\dagger}{\cal O}-\Sigma^{\prime\dagger}\Sigma')\nonumber\\
&&\times [{\rm det}({\cal O}^{\dagger}{\cal O})]^2\theta(\frac{{\rm
det}{\cal O}}{\sqrt{{\rm det}[{\cal O}^{\dagger}{\cal O}]}})
{\cal F}[{\cal O}]\delta(\Sigma'-\Sigma^{\prime\dagger})\nonumber\\
&=&\frac{1}{2}{\rm det} {\cal O}~
\delta(1-\frac{{\rm det}{\cal O}^{\dagger}}{{\rm det}{\cal O}})
~{\cal F}[{\cal O}]~\nonumber\\
&&\times\int{\cal D}\Sigma'\;
\delta({\cal O}^{\dagger}{\cal O}-\Sigma^{\prime\dagger}\Sigma')
\delta(\Sigma'-\Sigma^{\prime\dagger})\nonumber\\
&=&\frac{1}{2}{\rm det} {\cal O}~\delta(0)~
{\cal F}[{\cal O}]~\nonumber\\
&&\times\int{\cal D}\Sigma'\;
\delta({\cal O}^{\dagger}{\cal O}-\Sigma^{\prime\dagger}\Sigma')
\delta(\Sigma'-\Sigma^{\prime\dagger}).
\end{eqnarray}
In the last step, we have used the property ${\rm det}{\cal O}={\rm
det}{\cal O}^\dagger$. Taking ${\cal F}[{\cal O}]$ to be
\begin{eqnarray}                     
&&\frac{1}{{\cal F}[{\cal O}]}\equiv{\rm det} {\cal O}~
\int{\cal D}\Sigma'\;
\delta({\cal O}^{\dagger}{\cal O}-\Sigma'^{\dagger}\Sigma')
\delta(\Sigma'-\Sigma'^{\dagger}),\label{F'}
\end{eqnarray}
eq. (A3) becomes
\begin{eqnarray}                      
&&\int{\cal D}U\;\delta(U^{\dagger}U-1)\delta({\rm det}U-1)\nonumber\\
&&\times{\cal F}[{\cal O}]~\delta(\Omega {\cal O}^{\dagger}\Omega
-\Omega^{\dagger} {\cal O}\Omega^{\dagger})\nonumber\\
&&={\rm const},
\end{eqnarray}
which is of the form of eq.(\ref{Uin}) in the text.

Next we look at the meaning of the variable $\Sigma'$ in (A4). The constraints
on $\Sigma'$ in (A4) are
\begin{eqnarray}                    
\Sigma'^\dagger=\Sigma',\hspace{1cm}\Sigma'^2={\cal O}^\dagger{\cal O}.
\end{eqnarray}
On the other hand, eqs.(\ref{Udef0}), (\ref{sigma}) and (\ref{O}) in
the text show that the $\sigma$-field is constrained as
\begin{eqnarray}                    
\sigma^\dagger=\sigma,
\hspace{1cm}(\Omega^\dagger\sigma\Omega)^2={\cal O}^\dagger{\cal O}.
\end{eqnarray}
Comparing (A6) with (A7), we find $\Sigma'\sim \Omega^\dagger\sigma\Omega$. We 
know that the definition of $\sigma$ is not unique. It is up to a hidden 
symmetry transformation $\sigma\to h^\dagger\sigma h$. Therefore $\Sigma'$ can 
be regarded as an equivalent definition of $\sigma$, and thus (A4) can be 
written as 
\begin{eqnarray}                   
&&\frac{1}{{\cal F}[{\cal O}]}={\rm det} {\cal O}~
\int{\cal D}\sigma\;
\delta({\cal O}^{\dagger}{\cal O}-\sigma^{\dagger}\sigma)
\delta(\sigma-\sigma^{\dagger}),\nonumber\\
&&
\end{eqnarray}
which is just eq.(\ref{F}) in the text.

\end{document}